\documentclass{aa}  
\usepackage{lscape}
\usepackage{graphicx}
\usepackage{txfonts}
\usepackage{comment} 
\usepackage{color} 
\usepackage{amsmath} 
\usepackage{ulem} 
\usepackage{url}
\usepackage{orcidlink}

\newcommand{\Msun}{\mbox{M$_{\odot}$}}
\newcommand{\Mstar}{\mbox{M$_{\star}$}}
\newcommand{\Rstar}{\mbox{R$_{\star}$}}
\newcommand{\Lstar}{\mbox{L$_{\star}$}}
\newcommand{\Teff}{\mbox{T$^{\text{eff}}_\star$}}

\newcommand{\Rsun}{\mbox{R$_{\odot}$}}

\defcitealias{Kunitomo2021}{K21}
\begin{document} 

 \title{Planet formation around Intermediate-mass stars I:}
\subtitle{Different disc evolutionary pathways as a function of stellar mass}

 \titlerunning{Planet formation around intermediate-mass stars}

\author{M. P. Ronco\orcidlink{0000-0003-1385-0373}\inst{1,5},
M. R. Schreiber \orcidlink{0000-0003-3903-8009}\inst{2,5},
E. Villaver \orcidlink{0000-0003-4936-9418}\inst{3,4},
O. M. Guilera \orcidlink{0000-0001-8577-9532}\inst{1,5},
M.M. Miller Bertolami \orcidlink{0000-0001-8031-1957}\inst{1}
}
\authorrunning{Ronco et al. }
\offprints{M. P. Ronco\\ \email{mpronco@fcaglp.unlp.edu.ar}}
\institute{Instituto de Astrof\'{\i}sica de La Plata, CCT La Plata-CONICET-UNLP, Paseo del Bosque S/N (1900), La Plata, Argentina, 
\and
{Departamento de F{\'i}sica, Universidad T\'ecnica Federico Santa Mar\'ia, Av. España 1680, Valpara{\'i}so, Chile.}
\and
Instituto de Astrof{\'i}sica de Canarias, E-38205 La Laguna, Tenerife, Spain.
\and
Agencia Espacial Espa\~{n}ola (AEE), 41015 Sevilla, Spain.
\and
N\'ucleo Milenio de Formaci\'on Planetaria (NPF), Chile.
\\
}

   \date{Received ; accepted }

 
  \abstract
  { 
 The study of protoplanetary disc evolution and theories of planet formation has predominantly concentrated on solar (and low) mass stars since they host the majority of the confirmed exoplanets. Nevertheless, the confirmation of numerous planets orbiting stars more massive than the Sun (up to $\sim$ 3\,\Msun) has sparked considerable interest in understanding the mechanisms involved in their formation, and thus, in the evolution of their hosting protoplanetary discs. }
   {
We aim at improving our knowledge on the evolution of the gaseous component of protoplanetary discs around intermediate mass stars and to set the stage for future studies of planet formation around them.
}
   {We study the long-term evolution of protoplanetary discs affected by viscous accretion and photoevaporation by X-ray and FUV photons from the central star around stars in the range of 1 - 3\Msun~considering the effects of stellar evolution and solving the vertical structure equations of the disc. We explore the effect of different values of the viscosity parameter and the initial mass of the disc.
   }
    {We find that the evolutionary pathway of protoplanetary disc dispersal due to photoevaporation depends on the stellar mass. Our simulations reveal four distinct evolutionary pathways for the gas component not reported before that are a consequence of stellar evolution, and which likely have a substancial impact on the dust evolution and thus on planet formation. 
    As the stellar mass increases from one solar mass to $\sim$1.5 - 2\Msun, the evolution of the disc changes from the conventional inside-out clearing, in which X-ray photoevaporation generates inner holes, to a homogeneous disc evolution scenario where both inner and outer discs, formed after the gap is opened by photoevaporation, vanish over a similar timescale.
    As the stellar mass continues to increase, reaching $\sim$2 - 3\Msun, we have identified a distinct pathway that we refer to as revenant disc evolution. In this scenario, the inner and outer discs reconnect after the gap opened. For the largest masses, we observe outside-in disc dispersal, in which the outer disc dissipates first due to a stronger FUV photoevaporation rate. 
   Revenant disc evolution stands out as it is capable of extending the disc lifespan. 
    Otherwise, the disc dispersal time scale decreases with increasing stellar mass except for low viscosity discs. 
    } 
   {}
   \maketitle
%
\section{Introduction}
The discovery of several thousand extra-solar planets in the last decades \citep{Zhu2021} shows that planet formation is a common and robust process, at least around low-mass stars and sun-like stars. The large amount of discovered planetary systems, and the variety of these in terms of number of planets, planet masses, orbital separations and, more general the overall architecture of planetary systems, makes planet formation one of the most fascinating topics of modern astrophysics.  
Understanding the planet formation process is inherently related to understanding the evolution of protoplanetary discs.

Planets grow by the concurrent accretion of solids and gas (see the recent reviews of \citealt{Venturini2020Rev} and \citealt{Drazkowska22}). According to the core-accretion scenario, gas giant planets form from a core built by the accretion of pebbles \citep{Ormel10,Lambrechts12,Lambrechts14,JohansenLambrechts2017,Lambrechts19}, planetesimals \citep{Mordasini2009,Ronco2017} or both \citep{Alibert2018,Guilera2020}, that grows sufficiently to efficiently accrete gas from the surrounding disc \citep{Pollack1996, Ikoma2000, Alibert05,Guilera2010,Venturini16,Guilera2020, Ormel2021}. Consequently, the time scale available for gas giant planets to form
is largely determined by the evolution of the gas component of protoplanetary discs \citep{Ronco2017,Guilera2020,Venturini2020SE,Venturini2020Letter}.  
As soon as the mass supply from the surrounding cloud becomes negligible, viscous accretion slowly decreases the mass and the accretion rate of 
the disc \citep[e.g.][]{Pringle1981}. 
Ionizing radiation creates a layer of relatively hot ionized gas and beyond a certain radius, called the gravitational radius, the local thermal energy of the ionized gas exceeds its gravitational energy and the gas escapes in form of a wind.  
When the accretion rate becomes similar to the photoevaporation rate, a gap opens in the disc, and the inner disc drains on its 
viscous time scale. This causes the outer disc to be more efficiently irradiated by ionizing radiation which causes it to disperse fast \citep{Alexander2006a, Alexander2006b}.

Most studies of the evolution of protoplanetary discs focus on low-mass and solar mass stars and assume a constant ionizing flux emitted by the central star (see e.g. \citealt{Clarke2001,Alexander2006a,Gorti2009, Owen2010,Owen2012,ErcolanoPascucci2017}). 
In fact, \citet{Gorti2009} was the first to include 
a time-dependent FUV luminosity combined with constant X-ray and EUV luminosities in models of protoplanetary disc evolution. More recently, \citet{Kunitomo2021}, from now on 
\citetalias{Kunitomo2021}, extended the early work by \citet{Gorti2009} by incorporating 
analytical prescriptions for all types of ionizing radiation, from the FUV to the X-ray regime. In addition, \citetalias{Kunitomo2021} investigated not only the evolution of discs around sun-like stars 
but also studied the long-term evolution of protoplanetary discs affected by photoevaporation around intermediate mass stars, covering the mass range of $0.5-5$\,\Msun~and considering stellar evolution. These authors also showed that for low-mass stars ($\text{M}_{\star} \lesssim 2.5\text{M}_{\odot}$) photoevaporation is mainly driven by X-ray irradiation, while for intermediate-mass stars ($2.5\Msun \lesssim \Mstar \lesssim 5\Msun$) photoevaporation is dominated by FUV irradiation.

Studies of discs provide invaluable clues to the requisite conditions essential to planetary system formation. Improving our understanding of disc evolution and planet formation around intermediate mass stars is important for several reasons. 
Although the detection of planets around these stars is not yet as prolific as around their less massive counterparts,  planets have been detected through direct imaging around young A-stars, typically still embedded in a surrounding debris disc, the most famous examples being the planets around $\beta$\,Pic \citep{Lagrange2009, Lagrange2010,Lagrange2019} and HD\,8799 \citep{Marois2008,Marois2010}. Furthermore, the results of radial velocity campaigns  
targeting giant stars have been pretty successful (see e.g. \citealt{Nied2016a}), revealing a non-negligible occurrence rate of giant planets around 1 -- $3\,\Msun$ stars \citep{Nied2016b}
with a peak at $1.68\pm0.59$\,\Msun \citep{Wolthoff2022}. The dearth of stars with planetary companions
beyond this peak might be related to the shorter lifetimes of discs around stars more massive than $\sim2$\,\Msun~ \citep[e.g.][]{Ribas15}. Thus, studying disc evolution and planet formation around intermediate mass stars is just an important component of aiming at a better understanding of planet formation in general. 
In addition, observations of Herbig Ae/Be (HAeBe) stars, which are intermediate mass stars hosting protoplanetary discs, revealed disc structures that might be related to the planet formation process, star-disc interactions, and/or the accretion process \citep{Millan2007, Kraus2015}. Comparing model predictions with these observations may help to understand these complicated processes. Last but not least, the large number of metal polluted white dwarfs with masses below $0.7$\,\Msun~\citep[][their figure 1]{Koester2014} might imply that planet formation around stars in the mass range $1-3$\,\Msun~is potentially rather efficient.

While the study conducted by \citetalias{Kunitomo2021} represents a huge step forward in our comprehension of photoevaporation and the lifetimes of protoplanetary discs around intermediate mass stars, a detailed exploration of the input parameters is still lacking. As a first step towards a better understanding of disc evolution and planet formation around these stars, we here extend their work by calculating a finer grid of evolutionary sequences 
of protoplanetary discs around $1-3$\,\Msun~stars for different values of the viscosity parameter $\alpha$ and the initial mass of the disc considering stellar evolution and fully solving the vertical structure equations. 

We find four fundamentally different evolutionary pathways for the evolution of the gas component that are direct consequence of the photoevaporation process afected by stellar evolution that were not reported before and that likely have a significant impact on the evolution of the dust and thus on planet formation scenarios. 

\section{General Description of our model}
 
The results presented in this work have been calculated with {\scriptsize PLANETALP}, a 1D+1D model describing the evolution of the gas and the dust component of protoplanetary discs as well as the planet formation process \citep{Ronco2017,Guilera2017Letter,Guilera2019,Guilera2020,Guilera2021}. Unlike the code used in \citetalias{Kunitomo2021} and previous works \citep{Suzuki2016,Kunitomo2020}, {\scriptsize PLANETALP} fully solves the vertical structure equations for each radial bin as described in detail in \citet{Guilera2017Letter, Guilera2019}. 

The main goal of this paper is to set the stage for simulations of the planet formation process around stars more massive than the Sun. In this work we focus entirely on the evolution of the gas disc. The time evolution of the solid component and the impact of the evolution of both components on the process of planet formation will be addressed in a subsequent paper. 

\subsection{Vertical structure and Gas disc evolution}

To solve the vertical structure of the disc we assume an axisymmetric, thin, irradiated disc in hydrostatic equilibrium. We follow \citet{Guilera2017Letter,Guilera2019} who used the methodology described in \citet{Alibert05} and \citet{Migaszewski2015}, and solve for each radial bin the disc structure equations given by,
\begin{eqnarray}
  \frac{\partial P}{\partial z} = -\rho \Omega^2 z, 
    ~~~{\frac{\partial F}{\partial z}} = \frac{9}{4} \rho \nu \Omega^2,~~~
    \frac{\partial T}{\partial z} = \nabla \frac{T}{P}\frac{\partial P}{\partial z},
 \label{eq:Temperatura} 
\end{eqnarray}
where $P$, $\rho$, $F$, $T$ and $z$ represent the pressure, density, radiative heat flux, temperature and vertical coordinate of the disc, respectively. $\Omega$ is the Keplerian frequency at a given radial distance, while $\nu=\alpha c^{2}_{\text{s}}\Omega$ is the kinematic viscosity with $\alpha$ a dimensionless parameter \citep{ShakuraSunyaev1973}, and $c^{2}_{\text{s}}$ the square of the locally isothermal speed of the sound.

The mechanisms that heat the disc are viscosity and irradiation from the central star. This energy is vertically transported by radiation and convection according to the standard Schwarzschild criterion. The central object is assumed to be a protostar of mass \Mstar, radius \Rstar~ and effective temperature \Teff. The latter two parameters determine the irradiation of the disc's surface by the central star, given by
\begin{eqnarray}
  T_{\text{irr}} &=& \Teff\left[ \frac{2}{3\pi} \left( \frac{\Rstar}{R} \right)^3 + \frac{1}{2} \left(\frac{\Rstar}{R} \right)^2 \left( \frac{H}{R} \right)
  \left( \frac{d\log H}{d\log R} - 1 \right) \right]^{0.5}, \label{eq:temp_irrad} 
\end{eqnarray}
where $R$ is the is the radial coordinate, and $d\log H/d\log R = 9/7$ \citep{Chiang1997} For different stellar masses we adopt the values 
for \Rstar~and \Teff~that were used by \citetalias{Kunitomo2021} as initial conditions to compute the evolutionary tracks with \url{MESA} \citep{Paxton2011}. In other words, we use values of \Rstar~and \Teff~corresponding to the birthline ($t=0$) of \citet{StahlerPalla2004}. 
While it is clear that \Rstar~and \Teff~change with time due to the stellar evolution, we are not considering the time evolution of these values to compute the vertical structure of the disc. We discuss the impact of this simplification in Appendix B. 

Once the vertical structure is computed, the averaged gas surface density and viscosity are used to solve the classical radial diffusion equation \citet{Pringle1981}:
\begin{align}
  \frac{\partial \Sigma_{\text{g}}}{\partial t}= & \frac{3}{R}\frac{\partial}{\partial R} \left[ R^{1/2} \frac{\partial}{\partial R} \left( \overline{\nu} \Sigma_{\text{g}} R^{1/2} \right) \right] - \dot{\Sigma}_{\text{PE}}(R)
\label{eq:evol_gas}
\end{align}
which describes the time evolution of the gaseous disc driven by viscous accretion and photoevaporation caused by high energy photons emitted by the central star. In this equation, which is solved considering zero torque as boundary conditions (imposing zero density), $\Sigma_{\text{g}}$ is the gas surface density, and $\dot{\Sigma}_{\text{PE}}$ the photoevaporation rate. 

\subsection{X-ray and FUV photoevaporation rates}
\label{sec:photoevaporation_rates}

Knowing the time evolution of the high-energy irradiation is fundamental to compute the time dependence of the X-ray and FUV photoevaporation rates. 

In contrast to our previous works
\citep{Ronco2017,Guilera2020,Venturini2020SE,Venturini2020Letter, Ronco2021, Guilera2021}, we here compute the evolution of the gas disc by considering the evolution of the main stellar parameters, i.e. luminosity \Lstar~, effective temperature \Teff, stellar radius \Rstar, X-ray luminosity $L_\text{X}$ and the FUV luminosity $L_{\text{FUV}}$, following Table 1 of \citetalias{Kunitomo2021}.

As in \citetalias{Kunitomo2021}, mass loss due to photoevaporation $\dot{\Sigma}_{\text{PE}}(R)$ in our model takes only irradiation by the central star into account and is computed as
\begin{align}
  \dot{\Sigma}_{\text{PE}}(R) = \text{max} [\dot{\Sigma}_{\text{X-ray}}(R),\dot{\Sigma}_{\text{FUV}}(R)],
\label{eq:photo}
\end{align}
where $\dot{\Sigma}_{\text{X-ray}}(R)$ and $\dot{\Sigma}_{\text{FUV}}(R)$ represent the X-ray and FUV photoevaporation rates, respectively. In this work we ignore EUV photoevaporation as it is negligible for stellar masses between $1\Msun$ and $3\Msun$ \citepalias[see][their figure 11]{Kunitomo2021}, which are the ones of interest here. 
For the X-ray photoevaporation rate we follow \citetalias{Kunitomo2021} and consider a simplified version of the prescription proposed by \citet{Owen2012}, that is   
\begin{equation}
\dot{\Sigma}_{\text{X-ray}}(R) = \dot{\Sigma}_{\text{X,0}}\left(\dfrac{L_\text{X}}{10^{30}\text{erg s}^{-1}}\right)\left(\dfrac{R}{2.5~\text{au}}\right)^{-2}
\label{eq:XRrate}
\end{equation}
for primordial discs beyond 2.5~au~$(\Mstar/\Msun)$, where $\dot{\Sigma}_{\text{X,0}}=5.1\times10^{-12}$ g s$^{-1}$ cm$^{-2}$. The gas confined within the inner region of the disc is gravitationally bound and is not affected by photoevaporation, thus $\dot{\Sigma}_{\text{X-ray}}(R)=0$. For discs with holes that suffer direct photoevaporation we follow \citet[][their Appendix B (B2)]{Owen2012}. 

The FUV photoevaporation rate is computed as 
\begin{equation}
\dot{\Sigma}_{\text{FUV}}(R) = \dot{\Sigma}_{\text{FUV,0}}\left(\dfrac{L_{\text{FUV}}}{10^{31.7}\text{erg s}^{-1}}\right)\left(\dfrac{R}{4~\text{au}}\right)^{-2}
\label{eq:FUVrate}
\end{equation}
beyond 4~au~$(\Mstar/\Msun)$. Here $\dot{\Sigma}_{\text{FUV,0}}=10^{-12}$ g s$^{-1}$ cm$^{-2}$ and $L_{\text{FUV}}$, which is the FUV luminosity of the star, is defined as $L_{\mathrm{FUV}}=L_{\mathrm{FUV,acc}}+L_{\mathrm{FUV,ph}}+L_{\mathrm{FUV,chr}}$ \citep{Gorti2009,Kunitomo2021} with $L_{\mathrm{FUV,ph}}$ and $L_{\mathrm{FUV,chr}}$ being the photospheric and chromospheric FUV luminosities, respectively, and $L_{\mathrm{FUV,acc}}$ representing the FUV flux produced by accretion. Inside 4 au$(\Mstar/\Msun)$, $\dot{\Sigma}_{\text{FUV}}(R) = 0$. This prescription has been derived by \citetalias{Kunitomo2021} based on the results of \citet{Gorti2009} and \citet{WangGoodman2017}. For full details on the dependence of $L_{\text{X}}$ and $L_{\text{FUV}}$ on the stellar parameters, such as \Mstar, \Rstar, \Lstar~and \Teff, we refer the readers to sect. 2 in \citetalias{Kunitomo2021}.

In Appendix A we compare our model predictions with those presented in \citetalias{Kunitomo2021} finding very good agreement in terms of the evolution of the mass accretion and photoevaporation rates and disc dispersal timescales. 

\subsection{Initial conditions and scenarios to test}
 
To understand in more detail the impact of time-dependent photoevaporation on the evolution of the gas component 
of protoplanetary discs, we defined a grid of initial disc and stellar parameters.  
For the mass of the central star we used $\Mstar = 1\Msun,~ 1.5\Msun,~2\Msun,~2.5\Msun$ and $3\Msun$. 
For the initial gas surface density profile we adopt, as in \citetalias{Kunitomo2021}, the steady state solution \citep{Lynden-Bell1974} given by:
\begin{eqnarray}
  \Sigma_{\text{g}}(R, t=0) &=& \dfrac{M^0_{\text{d}}}{2\pi R^2_\text{c}} \dfrac{e^{-(R/R_\text{c})}}{R/R_\text{c}},\label{eq1-sec2-0}
\end{eqnarray}
where $M^0_{\text{d}}$ is the initial mass of the disc and $R_\text{c}$ the characteristic or cut-off radius. 

Following \citet{Andrews2010} we assumed a linear relation between the initial disc mass, ${M}_{\text{d}}$, and the stellar mass, i.e.,
\begin{eqnarray}
\text{M}_{\text{d}} &=& \text{M}_{\text{d}}^0 \, \left( \dfrac{\text{M}_{\star}}{\text{M}_{\odot}} \right),
\end{eqnarray}
considering for each stellar mass initially low-mass, intermediate, and massive discs, that is, we calculated the disc evolution for $\text{M}^0_{\text{d}}=0.03\Msun$, $0.06\Msun$ and $0.1\Msun$. 
 
With these initial masses,  all discs considered in this work are stable against gravitational collapse at every time step.
For the characteristic radius of the disc, $R_{\text{c}}$, we adopted the 
relation used in \citet{Burn2021},
\begin{eqnarray}
R_{\text{c}} &=& R^0_{\text{c}} \left( \dfrac{\text{M}_{\star}}{\text{M}_{\odot}} \right)^{0.625}, 
\end{eqnarray}
with $R^0_{\text{c}}=40$\,au, which corresponds to the mean value of the distributions inferred from observations \citep{Andrews2010}. 

As in \citet{Burn2021} we also scaled the inner edges of the discs $\text{R}_{\text{in}}$ using
\begin{eqnarray}
\text{R}_{\text{in}} &=& 0.1 ~\text{au}~\left( \dfrac{\text{M}_{\star}}{\text{M}_{\odot}} \right)^{1/3}.
\label{eq:inner-radius}
\end{eqnarray} 

For the viscosity we used, as in \citetalias{Kunitomo2021}, the following scaling law, 
\begin{eqnarray}
\alpha &=& \alpha_0 \left( \dfrac{\text{M}_{\star}}{\text{M}_{\odot}} \right),
\end{eqnarray}
with three different values for $\alpha_0$; $\alpha_0=10^{-2}$ (high viscosity), $\alpha_0=10^{-3}$ (intermediate), and $\alpha_0 = 10^{-4}$ (low viscosity). A summary of all parameters used in our simulations can be found in Table \ref{Tab1}.

\begin{table*}
\caption{Star and disc initial parameters for all the performed simulations. $\Mstar$ denotes the mass of the star, $\Rstar$ and $\Teff$ refer to the radius and effective temperature of the star at the birthline. $R_{\text{int}}$ and $R_{\text{c}}$ represent the inner border and the characteristic radius of the disc, respectively. $\alpha$ is the viscosity parameter and $\text{M}_{\text{d}}$ the mass of the disc. The last column refers to $\tau$, the disc dissipation timescale, which is not an initial condition but a result of our simulations (see sec. \ref{sec:timescales}). The cases highlighted in bold face are referring to the revenant discs that will be described in sec. \ref{Sec:RebornDisks}}.
\begin{center}
  \begin{tabular}{|c|c|c|c|c|c|c|c|c|}
    \hline
       Sim & $\Mstar$ $[\Msun]$ & $\Rstar$ $[\Rsun]$ & $\Teff$ [K] & $R_{\text{int}}$ [au] & $R_{\text{c}}$ [au] & 
       $\alpha$ & $\text{M}_{\text{d}}$ [$\Msun$] & $\tau$ [Myr]\\
      \hline
       S1 & 1.0 & 4.600 & 4482 & 0.1 & 40 & $10^{-2}$ & 0.03 & 0.55 \\
       S2 & 1.0 & 4.600 & 4482 & 0.1 & 40 & $10^{-2}$ & 0.06 & 1.02 \\
       S3 & 1.0 & 4.600 & 4482 & 0.1 & 40 & $10^{-2}$ & 0.10 & 1.61 \\
       S4 & 1.0 & 4.600 & 4482 & 0.1 & 40 & $10^{-3}$ & 0.03 & 1.19 \\
       S5 & 1.0 & 4.600 & 4482 & 0.1 & 40 & $10^{-3}$ & 0.06 & 2.80 \\
       S6 & 1.0 & 4.600 & 4482 & 0.1 & 40 & $10^{-3}$ & 0.10 & 7.00 \\
       S7 & 1.0 & 4.600 & 4482 & 0.1 & 40 & $10^{-4}$ & 0.03 & 3.49 \\
       S8 & 1.0 & 4.600 & 4482 & 0.1 & 40 & $10^{-4}$ & 0.06 & 8.04 \\
       S9 & 1.0 & 4.600 & 4482 & 0.1 & 40 & $10^{-4}$ & 0.10 & > 10 \\
       \hline
       S10 & 1.5 & 4.105 & 4797 & 0.114 & 51.5 & $1.5x10^{-2}$ & 0.045 & 0.64 \\
       S11 & 1.5 & 4.105 & 4797 & 0.114 & 51.5 & $1.5x10^{-2}$ & 0.09 & 0.97 \\
       S12 & 1.5 & 4.105 & 4797 & 0.114 & 51.5 & $1.5x10^{-2}$ & 0.15 & 1.52 \\
       S13 & 1.5 & 4.105 & 4797 & 0.114 & 51.5 & $1.5x10^{-3}$ & 0.045 & 1.42 \\
       S14 & 1.5 & 4.105 & 4797 & 0.114 & 51.5 & $1.5x10^{-3}$ & 0.09 & 2.65 \\
       S15 & 1.5 & 4.105 & 4797 & 0.114 & 51.5 & $1.5x10^{-3}$ & 0.15 & 4.58 \\
       S16 & 1.5 & 4.105 & 4797 & 0.114 & 51.5 & $1.5x10^{-4}$ & 0.045 & 5.05 \\
       S17 & 1.5 & 4.105 & 4797 & 0.114 & 51.5 & $1.5x10^{-4}$ & 0.09 & 6.10 \\
       S18 & 1.5 & 4.105 & 4797 & 0.114 & 51.5 & $1.5x10^{-4}$ & 0.15 & 8.54 \\
       \hline  
       S19 & 2.0 & 3.882 & 4797 & 0.126 & 61.7 & $2x10^{-2}$ & 0.06 & 0.66 \\
       S20 & 2.0 & 3.882 & 4797 & 0.126 & 61.7 & $2x10^{-2}$ & 0.12 & 0.98 \\
       S21 & 2.0 & 3.882 & 4797 & 0.126 & 61.7 & $2x10^{-2}$ & 0.20 & 1.42 \\
       S22 & 2.0 & 3.882 & 4797 & 0.126 & 61.7 & $2x10^{-3}$ & 0.06 & 1.38 \\
       S23 & 2.0 & 3.882 & 4797 & 0.126 & 61.7 & $2x10^{-3}$ & 0.12 & 2.64 \\
       S24 & 2.0 & 3.882 & 4797 & 0.126 & 61.7 & $2x10^{-3}$ & 0.20 & 3.56 \\
       S25 & 2.0 & 3.882 & 4797 & 0.126 & 61.7 & $2x10^{-4}$ & 0.06 & 6.40 \\
       S26 & 2.0 & 3.882 & 4797 & 0.126 & 61.7 & $2x10^{-4}$ & 0.12 & 7.60 \\
       \textbf{S27} & \textbf{2.0} & \textbf{3.882} & \textbf{4797} & \textbf{0.126} & \textbf{61.7} & $\mathbf{2x10^{-4}}$ & \textbf{0.20} & \textbf{9.52}\\
       \hline
       S28 & 2.5 & 4.204 & 5178 & 0.136 & 70.9 & $2.5x10^{-2}$ & 0.075 & 0.57 \\
       S29 & 2.5 & 4.204 & 5178 & 0.136 & 70.9 & $2.5x10^{-2}$ & 0.15 & 0.84 \\
       S30 & 2.5 & 4.204 & 5178 & 0.136 & 70.9 & $2.5x10^{-2}$ & 0.25 & 1.14 \\
       S31 & 2.5 & 4.204 & 5178 & 0.136 & 70.9 & $2.5x10^{-3}$ & 0.075 & 1.14 \\
       S32 & 2.5 & 4.204 & 5178 & 0.136 & 70.9 & $2.5x10^{-3}$ & 0.15 & 2.51 \\
       \textbf{S33} & \textbf{2.5} & \textbf{4.204} & \textbf{5178} & \textbf{0.136} & \textbf{70.9} & $\mathbf{2.5x10^{-3}}$ & \textbf{0.25} & \textbf{3.65} \\
       S34 & 2.5 & 4.204 & 5178 & 0.136 & 70.9 & $2.5x10^{-4}$ & 0.075 & 8.50 \\
       \textbf{~~S35$^\star$}& \textbf{2.5} & \textbf{4.204} & \textbf{5178} & \textbf{0.136} & \textbf{70.9} & $\mathbf{2.5x10^{-4}}$ & \textbf{0.15} & \textbf{9.50} \\
       S36 & 2.5 & 4.204 & 5178 & 0.136 & 70.9 & $2.5x10^{-4}$ & 0.25 & > 10\\
       \hline 
       S37 & 3.0 & 4.571 & 5415 & 0.144 & 79.5 & $3x10^{-2}$ & 0.09 & 0.42 \\
       \textbf{~~S38$^\star$} & \textbf{3.0} & \textbf{4.571} & \textbf{5415} & \textbf{0.144} & \textbf{79.5} & $\mathbf{3x10^{-2}}$ & \textbf{0.18} & \textbf{1.15}\\
       S39 & 3.0 & 4.571 & 5415 & 0.144 & 79.5 & $3x10^{-2}$ & 0.30 & 1.23 \\
       \textbf{S40} & \textbf{3.0} & \textbf{4.571} & \textbf{5415} & \textbf{0.144} & \textbf{79.5} & $\mathbf{3x10^{-3}}$ & \textbf{0.09} & \textbf{1.70}\\
       S41 & 3.0 & 4.571 & 5415 & 0.144 & 79.5 & $3x10^{-3}$ & 0.18 & 2.17 \\
       S42 & 3.0 & 4.571 & 5415 & 0.144 & 79.5 & $3x10^{-3}$ & 0.30 & 2.25 \\
       S43 & 3.0 & 4.571 & 5415 & 0.144 & 79.5 & $3x10^{-4}$ & 0.09 & 9.75 \\
       S44 & 3.0 & 4.571 & 5415 & 0.144 & 79.5 & $3x10^{-4}$ & 0.18 & > 10 \\
       S45 & 3.0 & 4.571 & 5415 & 0.144 & 79.5 & $3x10^{-4}$ & 0.30 & > 10 \\      
      \hline
   \end{tabular}
  \end{center}
  \label{Tab1}
\end{table*}

\section{Results}

\begin{figure*}
   \centering
   \includegraphics[angle=0,width=1.\linewidth]{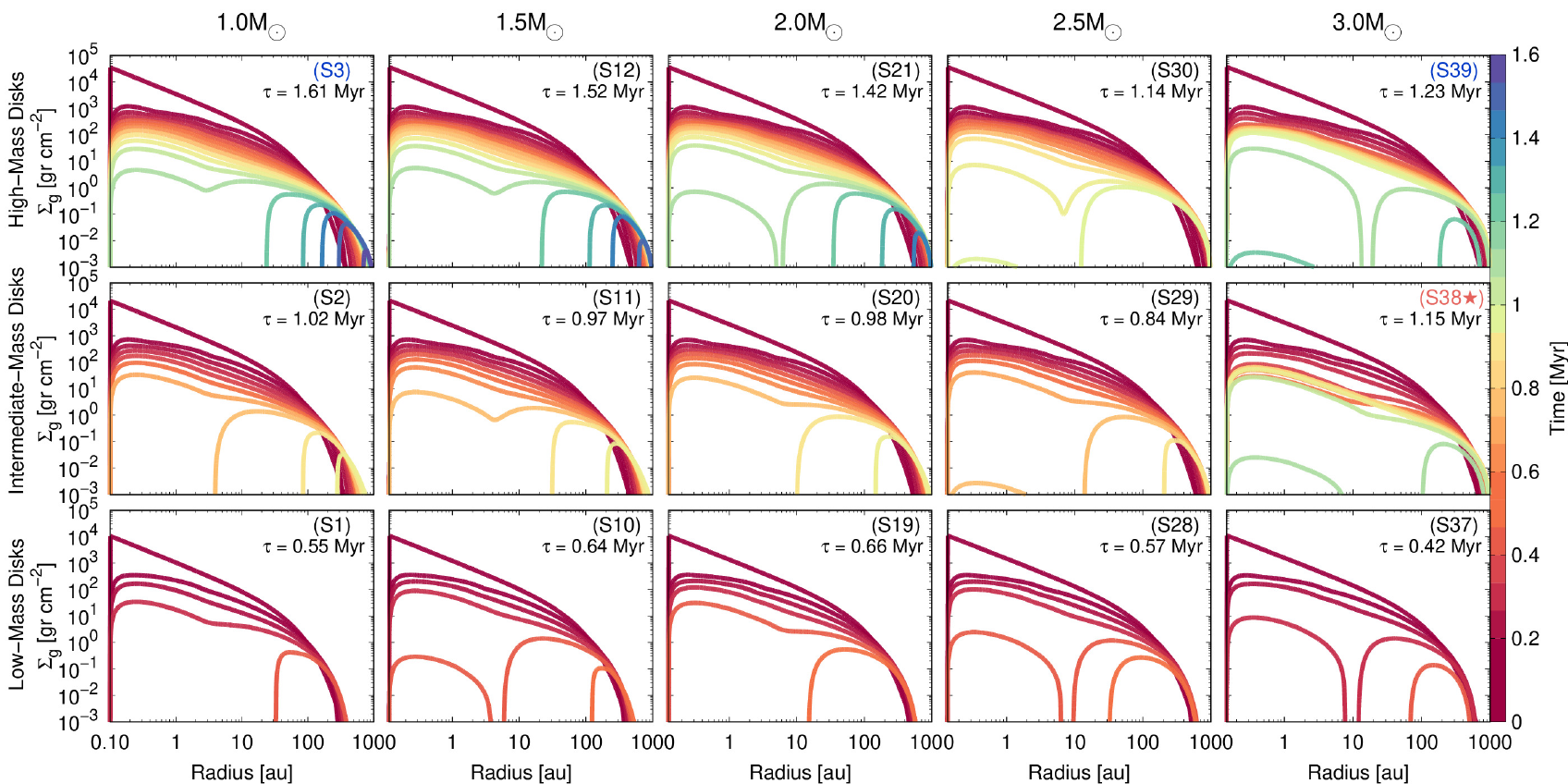}
   \includegraphics[angle=0,width=1.\linewidth]{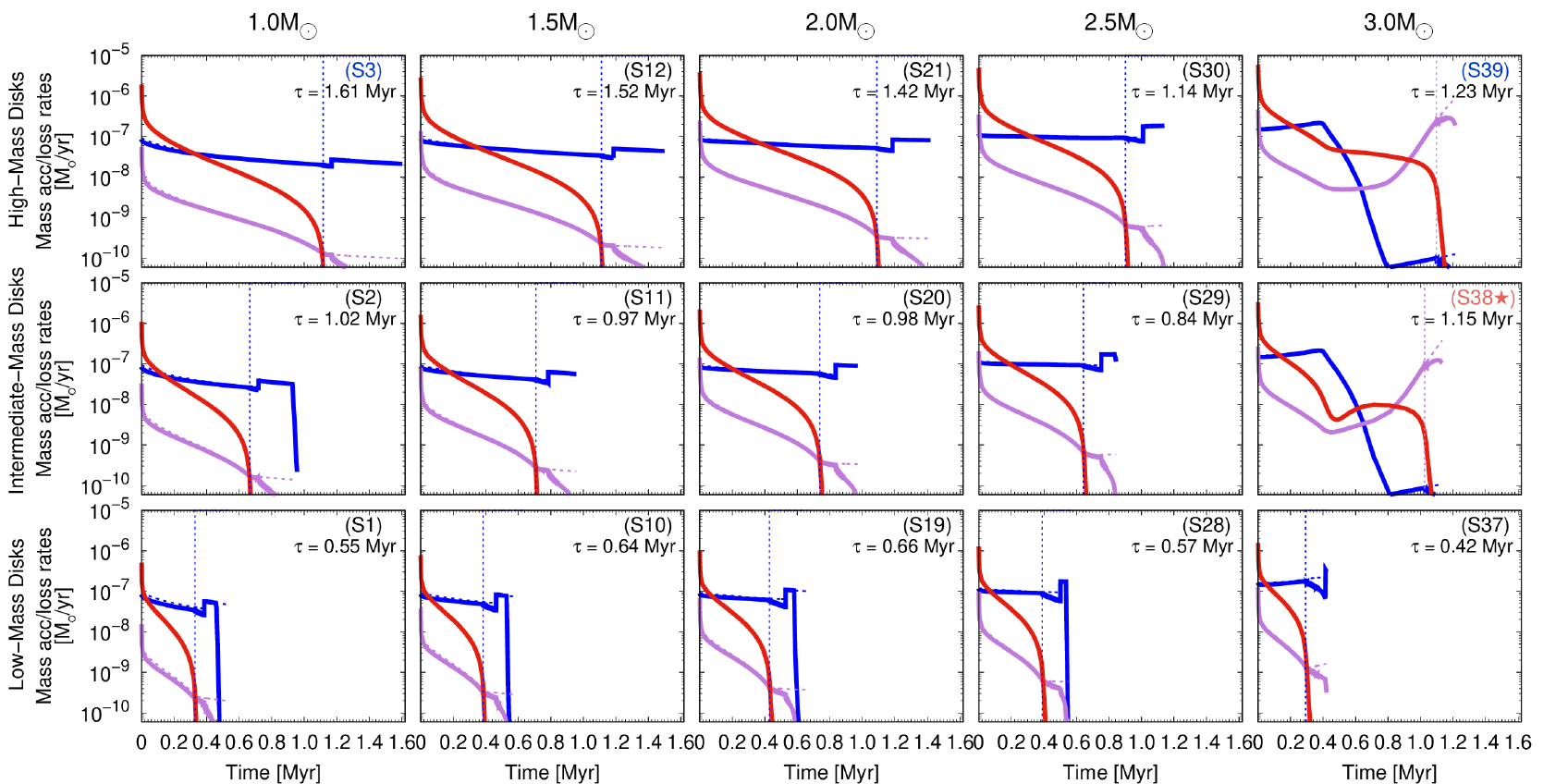}
   \caption{Upper panels: Time evolution of the gas surface density profiles for massive (top), intermediate (middle) and low (bottom) mass discs around stars with masses between $1\Msun$ and $3\Msun$ in discs with $\alpha_0=10^{-2}$, plotted every 0.1 Myr. The labels at the top right of each panel represent the simulation number and the dissipation timescale, also represented by the colorscale bar shown on the vertical right hand-side of the plot. The colored simulation numbers refer to specific cases described in the text. Lower Panels: Time evolution of the mass accretion rate $\dot{M}_{\text{acc}}$ (red solid curve) and the mass-loss rates $\dot{M_{\text{X}}}$ (blue solid curve) and $\dot{M_{\text{FUV}}}$ (lilac solid curve) by the X-ray and FUV photoevaporation, respectively, for the same discs, integrated along the regions where there is still gas surface density to remove. The dashed blue and lilac curves represent the same mass loss rates but integrated along the disc domain. The vertical dashed lines represent the time at which the gap opened in the disc, either by the effects of X-ray (blue) or by FUV (lilac) photoevaporation. For the high viscosity parameter assumed here most discs evolve following the classical inside-out disc dispersal (see Sect.\ref{Sec:inside-out}). Only for the highest stellar and disc masses (see cases S38 and S39) the outer and the inner disc disappear on similar time scales (a pathway that we call homogeneous disc evolution). 
   }
      \label{Fig:Perfiles-y-Mpunto-Alpha2}
    \end{figure*}

\begin{figure*}
   \centering
   \includegraphics[angle=0,width=1.\linewidth]{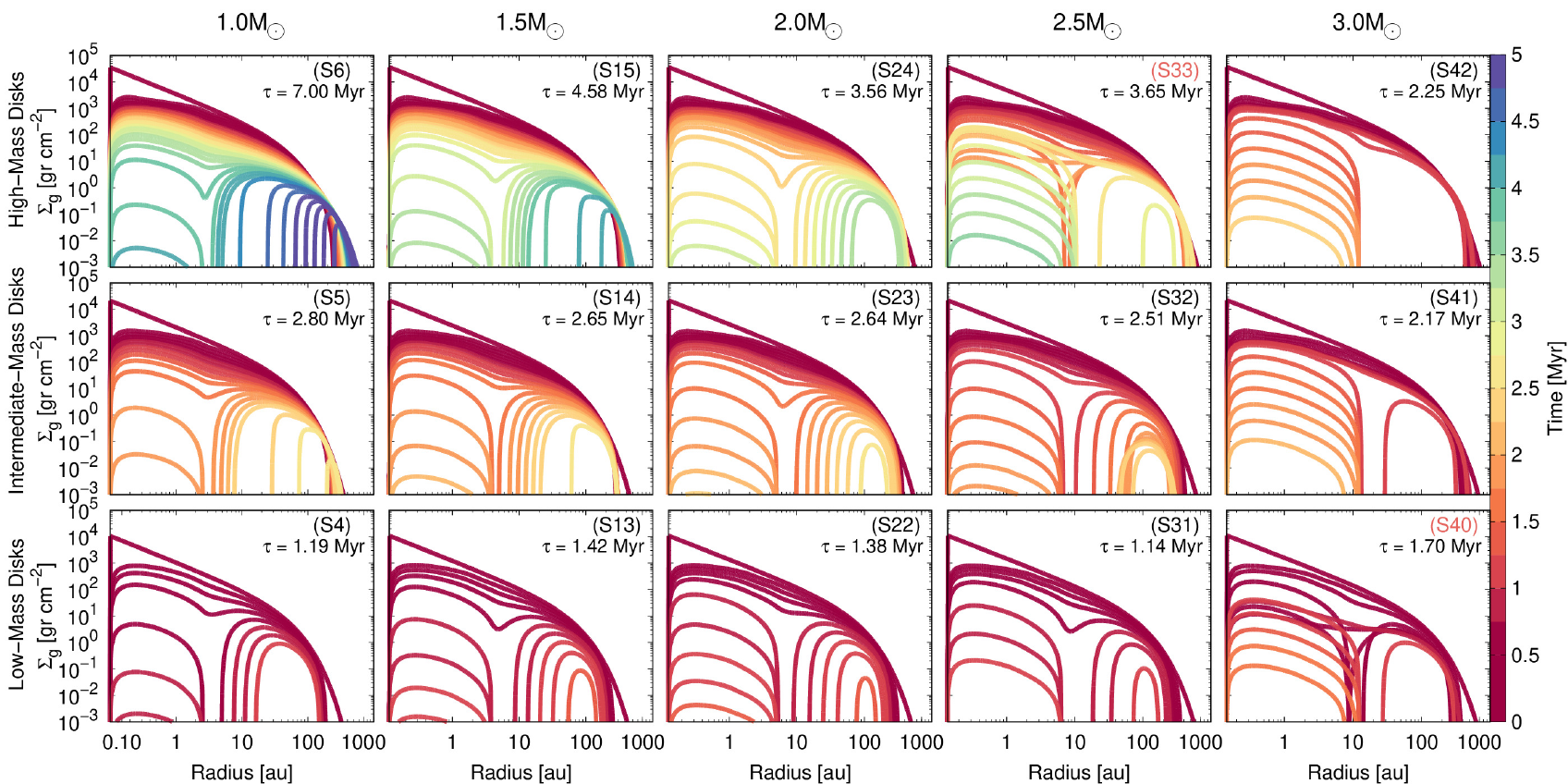}
   \includegraphics[angle=0,width=1.\linewidth]{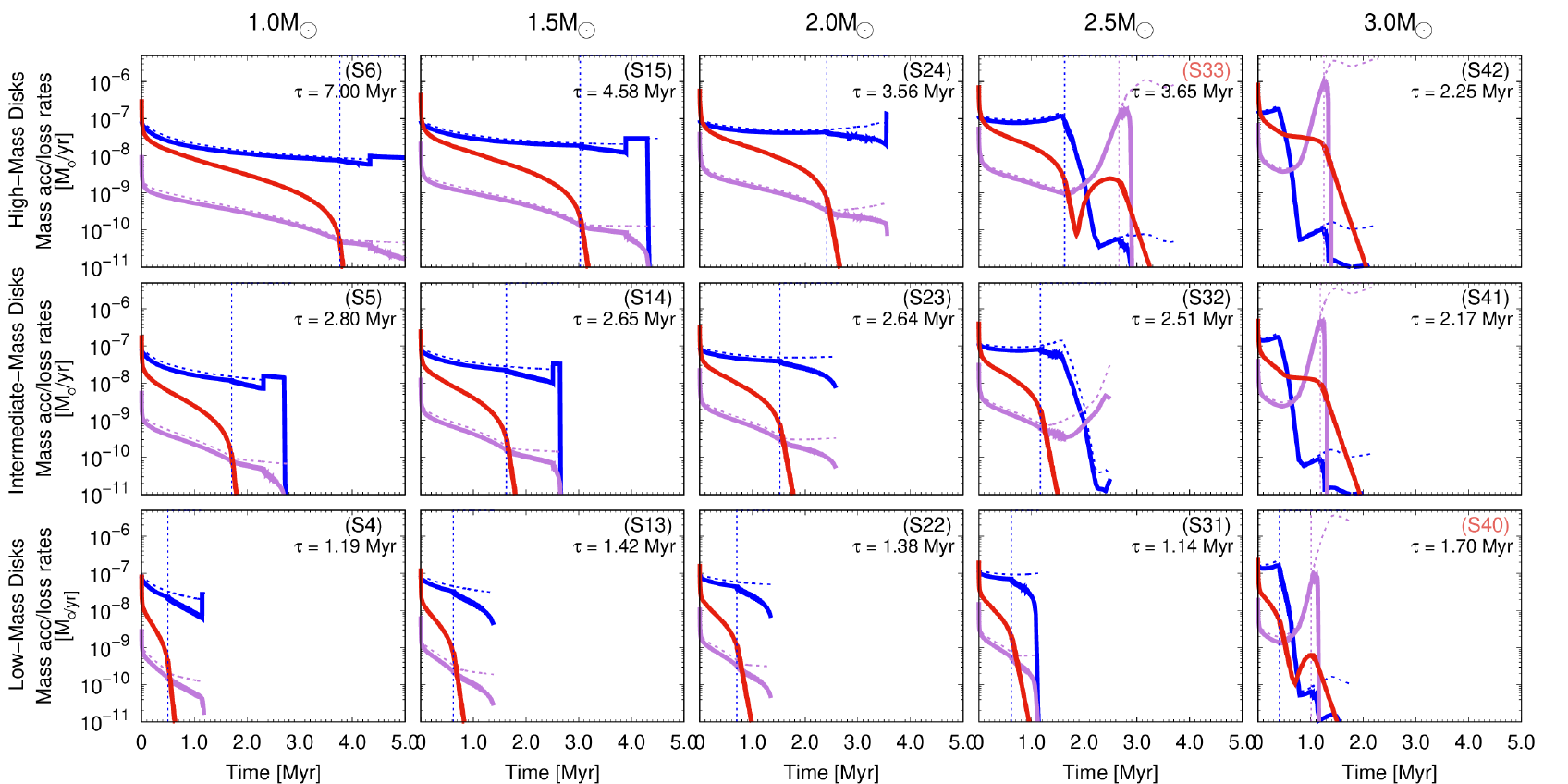}
  \caption{Same as fig. \ref{Fig:Perfiles-y-Mpunto-Alpha2} but for discs with $\alpha_0=10^{-3}$ and gas disc profiles plotted every 0.15 Myr. 
  For this intermediate value of the viscosity parameter we only predict classical inside-out disc dispersal for the most massive discs around the least massive stars (S5, S6, S15). 
  In the majority of cases, the discs dissipate according to homogeneous disc evolution, i.e. the inner and outer disc disappear nearly simultaneously (S4, S13, S14, S22, S23, S31, S32). For the initially more massive discs around the most massive stars (S41, S42) we predict the outer disc to disappear clearly before the inner disc, that is, in these cases the viscous time scale of the inner disc is longer than the photoevaporation time scale of the outer disc. 
  A particularly interesting evolutionary pathway is that predicted by S40 and S33. Following the gap opening through 
  X-ray photoevaporation, the outer and the inner disc reconnect 
  as the X-ray emission decreases and the final disc dispersal is caused by FUV photoevaporation. We termed this pathway revenant disc evolution (see Sect.\,\ref{Sec:RebornDisks} for details). 
  }
   \label{Fig:Perfiles-y-Mpunto-Alpha3}
\end{figure*}

\begin{figure*}
   \centering
\includegraphics[angle=0,width=1.\linewidth]{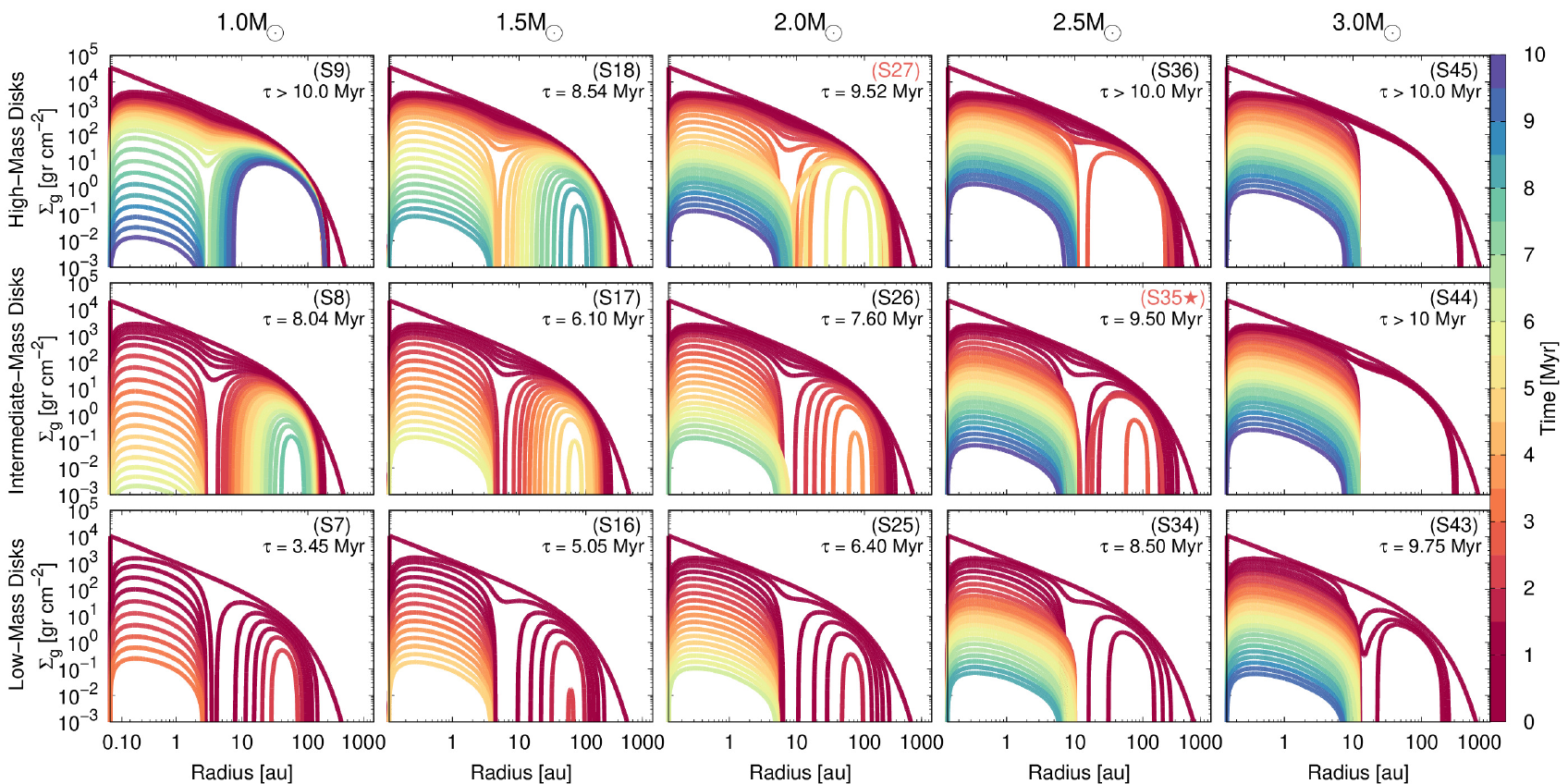}   
\includegraphics[angle=0,width=1.\linewidth]{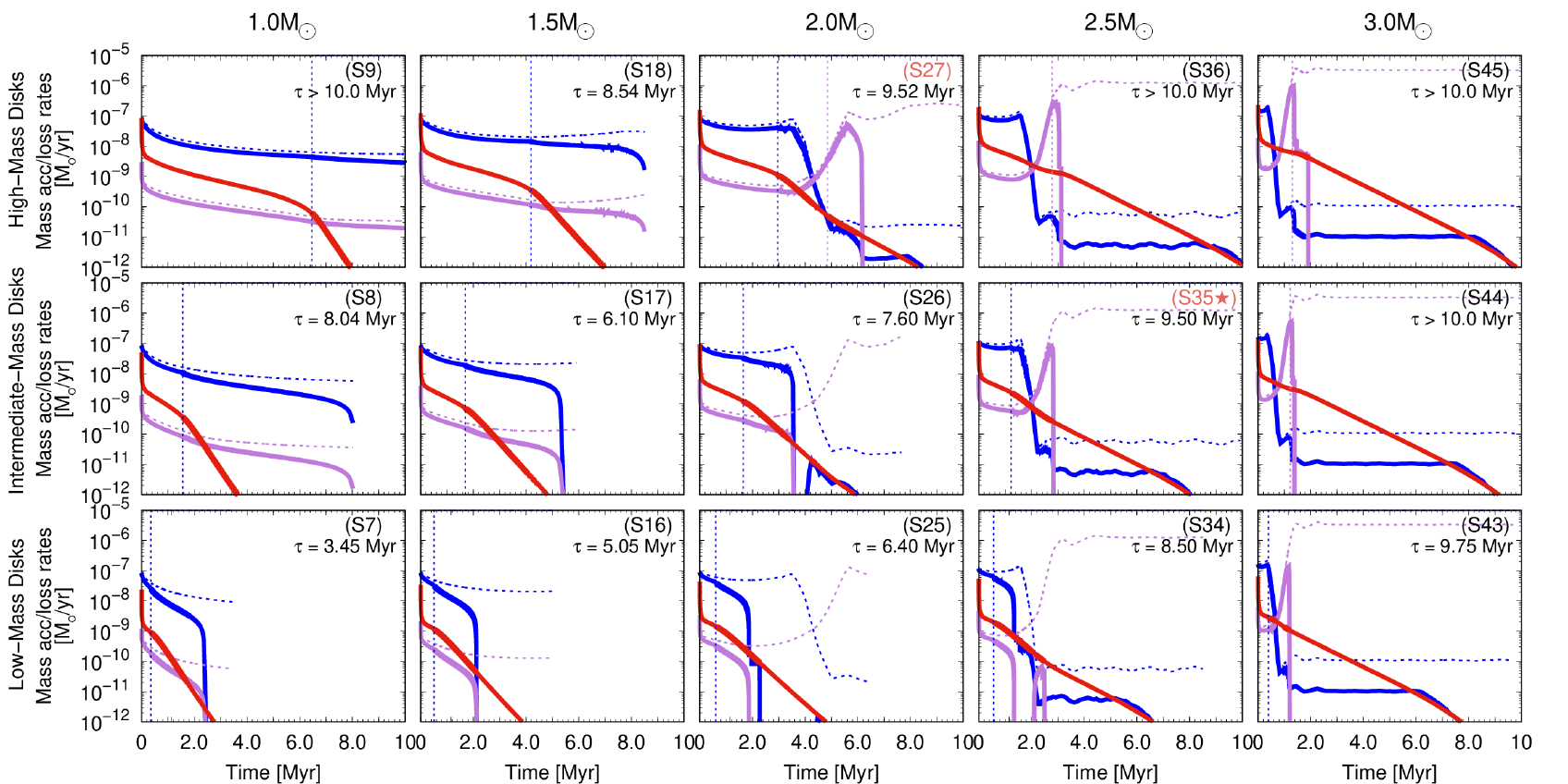}
  \caption{Same as figs. \ref{Fig:Perfiles-y-Mpunto-Alpha2} and \ref{Fig:Perfiles-y-Mpunto-Alpha3} but for discs with $\alpha_0=10^{-4}$ and gas disc profiles plotted every 0.35 Myr.
  For the lowest viscosity parameter in our grid, we did not find any disc that follows the classical inside-out disc dissipation pathway. Instead, for the lowest mass stars, we predict the photoevaporation time scale of the outer disc to be similar to the viscous time scale of the outer disc, which implies that both components disappear nearly simultaneously (homogeneous disc evolution, see Sect.\,\ref{sec:homogeneous}). With increasing stellar mass, the evolution switches first to revenant disc evolution (S27, S35*, see Sect.\,\ref{Sec:RebornDisks}) and then to outside-in disc dispersal (S34, S36, S43, S45, see Sect.\,\ref{Sec:outside-in} for details).   
  }
   \label{Fig:Perfiles-y-Mpunto-Alpha4}
\end{figure*}

In this section we describe in detail the results of our simulations. 
We illustrate the evolution of each disc by plotting the gas surface density as a function of radius and time for the three values for the viscosity parameter $\alpha_0$ that we adopted 
(see figs.\,\ref{Fig:Perfiles-y-Mpunto-Alpha2}, \ref{Fig:Perfiles-y-Mpunto-Alpha3} and \ref{Fig:Perfiles-y-Mpunto-Alpha4} for $\alpha_0 = 10^{-2}$, $10^{-3}$ and $10^{-4}$, respectively).
For each disc evolution we also show the time evolution of the accretion rate 
($\dot{M}_{\text{acc}}=-2\pi R\Sigma(R)v_{R}$, with $v_{R}$ the gas radial velocity), and the mass-loss rates due to X-ray ($\dot{M}_{\text{X}}=\int 2\pi R\dot{\Sigma}_{\text{X}}dR$) and FUV ($\dot{M}_{\text{FUV}}=\int 2\pi R\dot{\Sigma}_{\text{FUV}}dR$) photoevaporation integrated along the regions where there is still gas surface density to remove (bottom panels of the same figures). 
Moreover, for the sake of comparison with the work of \citetalias{Kunitomo2021} we also compute the total disc mass lost by photoevaporation as $M_{\text{PE}}=\int \dot{\Sigma}_{\text{PE}}dRdt$, integrated along the disc extension, where $\dot{\Sigma}_{\text{PE}}$ follows eq. \ref{eq:photo}. 

The main mass dissipation mechanisms for the vast majority of the discs we analysed are viscous evolution and X-ray photoevaporation (see bottom panels of figs.\,\ref{Fig:Perfiles-y-Mpunto-Alpha2}, \ref{Fig:Perfiles-y-Mpunto-Alpha3} and \ref{Fig:Perfiles-y-Mpunto-Alpha4}). 
FUV photoevaporation only becomes important for relatively massive discs around stars of $2.5- 3\,\Msun$. In these cases, it will have a strong impact on the evolution of the disc, as we will describe in detail in the following sections. 

Depending on the disc parameters and the stellar mass, we found four fundamentally different evolutionary pathways. 
We believe that the evolution of the gas component of protoplanetary discs we report here could significantly affect the planet formation process in these discs.

\subsection{Standard inside-out disc photoevaporation}
\label{Sec:inside-out}

The standard scenario for the disc dispersal of protoplanetary discs due to photoevaporation is frequently called inside-out disc dispersal. 
In this scenario, at first the disc evolves mainly due to viscous accretion until photoevaporation becomes effective and opens a gap at a few au separating an inner and an outer disc. 
Then, as the inner disc is orders of magnitude less massive than the outer one, it is rapidly accreted by the central star due to viscous accretion while the outer disc continues to loose a relatively small fraction of its mass due to photoevaporation. When the inner disc completely disappeared, the disc consists of a relatively massive outer disc and an inner cavity, i.e. it  
looks like a classical transition disc \citep{WilliamsCieza2011}.

As the outer disc is no longer protected from direct irradiation by the inner disc, the outer disc quickly photoevaporates. 

In our parameter study we find inside-out disc evolution in most of the simulations with high viscosity, particularly for intermediate and massive discs, and for very few cases with intermediate viscosity and intermediate/massive discs around Solar-type stars. Two examples of this behaviour are simulations S3 in fig. \ref{Fig:Perfiles-y-Mpunto-Alpha2} and S6 in fig. \ref{Fig:Perfiles-y-Mpunto-Alpha3}. 

\subsection{Homogeneous disc evolution}
\label{sec:homogeneous}

\begin{figure}
   \centering
   \includegraphics[angle=-90,width=0.99\columnwidth]{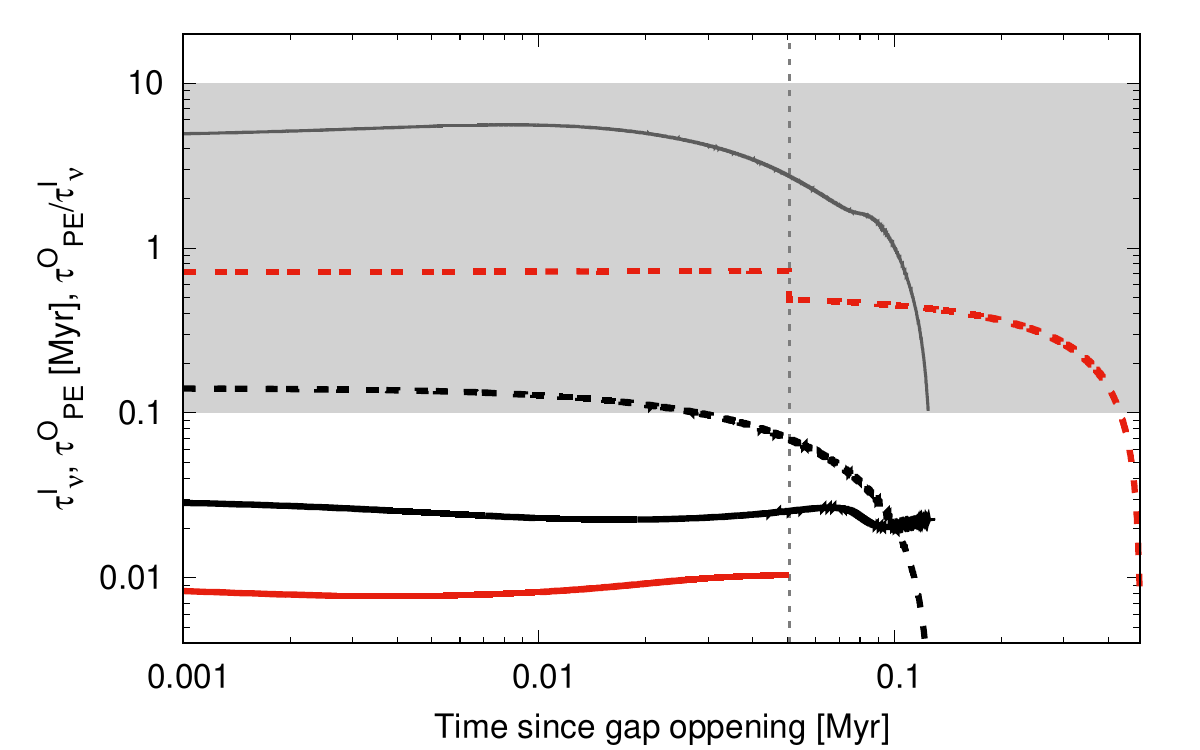}
   \caption{
   Viscous ($\tau^{\text{I}}_{\nu}$, solid lines) and photoevaporation ($\tau^{\text{O}}_{\text{PE}}$, dashed lines) time scales from the time of gap opening for the discs labelled S3 (red lines) and S39 (black lines). The dark gray solid curve represents the ratio $\tau^{\text{O}}_{\text{PE}}/\tau^{\text{I}}_{\nu}$ for the S39 disc that evolves always bounded within the gray shaded area, between 0.1 and 10. The same ratio for S3 falls completely outside the range of this figure. As a consequence and following our definition for homogeneous evolution, while model S39 describes the homogeneous disc evolution around a $3\Msun$ star (black lines), S3 represents inside-out evolution (red lines) of a disc around a star of $1\Msun$.  
The abrupt decrease in $\tau^{\text{O}}_{\text{PE}}$ exactly at the time the disc becomes a transition disc (vertical gray dashed line) is due to the change to the direct X-ray photoevaporation rate regime in S3. 
}
   \label{Fig:time-after-gap}
\end{figure}

In addition to the classical inside-out evolution generated by photoevaporation, we find different evolutionary pathways depending on the viscous and photoevaporation time scales after the gap opening. 
With increasing stellar mass, that is, moving from left to right in figs.\,\ref{Fig:Perfiles-y-Mpunto-Alpha2} and \ref{Fig:Perfiles-y-Mpunto-Alpha3}, the gas disc evolution becomes more 'homogeneous'. 
Following the formation of the gap, the inner and outer discs dissipate on comparable timescales, that is, the disc does not evolve through the configuration of an inner hole surrounded by a massive outer disc (classical transition disc).   

This change from inside-out to 'homogeneous' disc clearance is straight forward to understand.  
As the mass of the star increases, the gap opens further away from it (see eq. \ref{eq:XRrate}) which causes 
the inner disc to contain a larger mass fraction and to extend to larger radii. 
While the outer disc remains more massive by several orders of magnitude, the viscous time scale of the inner disc, that is $\tau^{\text{I}}_{\nu}=M_{\text{Inner}}/\dot{M}_{\text{acc}}$ (with $M_{\text{Inner}}$ the mass of the inner disc) becomes comparable to the photoevaporation timescale of the outer disc $\tau^{\text{O}}_{\text{PE}} = M_{\text{Outer}}/\dot{M}_{\text{PE}}$ (with $M_{\text{Outer}}$ the mass of the outer disc).  We define discs that follow 'homogeneous evolution' as those for which $0.1 \leq \tau^{\text{O}}_{\text{PE}}/\tau^{\text{I}}_{\nu} < 10$ after the gap opened 
for at least 95\% of the total time remaining until the complete dissipation of the disc.

In fig.\,\ref{Fig:time-after-gap} we compare the viscous ($\tau^{\text{I}}_{\nu}$, black lines) and photoevaporation  ($\tau^{\text{O}}_{\text{PE}}$, red lines) time scales for the discs labelled S3 and S39 (highlighted with blue simulation numbers S3 and S39 in fig. \ref{Fig:Perfiles-y-Mpunto-Alpha2}). While model S3 represents inside-out evolution (dashed lines) of a disc around a star of 1$\Msun$, S39 describes the homogeneous disc evolution around a $3\Msun$ star (solid lines). 
While for model S3 the difference between viscous and photoevaporation time scale corresponds throughout the evolution to roughly two orders of magnitude, for S39 the difference between both time scales is much smaller (blue solid curve) and remains bounded between 0.1 and 10.  
In fact, $\sim$ 0.1 Myr after gap opening, almost at the end of the disc's life time, the photoevaporation time scale of the outer disc becomes shorter than the viscous time scale of the inner disc which implies that in this case the outer disc disappears slightly before the inner disc. 

Among our simulations we find discs with a homogeneous evolution for all three viscosity parameters we considered. For $\alpha_0=10^{-2}$ we find homogeneous evolution around the most massive stars (2, 2.5 and 3$\Msun$ stars). For smaller viscosity parameters, we find that the transition from inside-out to homogeneous evolution moves to smaller masses. For $\alpha_0=10^{-3}$ 
and $\alpha_0=10^{-4}$ we find homogeneous evolution for stars with masses between 1.5 and 2.5$\Msun$ and 1 and $1.5\Msun$, respectively.  

The switch from inside-out to homogeneous disc evolution can also be seen in the bottom panels of 
figs.\,\ref{Fig:Perfiles-y-Mpunto-Alpha2}, \ref{Fig:Perfiles-y-Mpunto-Alpha3} and \ref{Fig:Perfiles-y-Mpunto-Alpha4}. 
If a given disc evolves through the transition disc stage, a drastic and immediate increase in $\dot{M}_{\text{X}}$ can be observed. This sudden change in $\dot{M}_{\text{X}}$ occurs if the inner disc has completely dissipated and does not shield the outer disc from the X-ray photons emitted by the star. With increasing stellar mass, the  
time during which the disc resembles a transition disc decreases until this stage is no longer reached
as the viscous time scale of the inner disc equals the photoevaporation time scale of the outer disc.  

\subsection{Outside-in disc evolution}
\label{Sec:outside-in}

We also found cases in our simulations where the order of the disc dispersal is completely reversed. 
In these cases, once the gap opens due to photoevaporation, the outer disc dissipates significantly faster than the inner one. 
A similar dispersal path has been found by \citet{Coleman2022} for discs around low-mass stars (0.1, 0.3 and 1$\Msun$). Note, however, that what causes this kind of evolution, referred by these authors as "Inside-Out with continued accretion" is a combination between X-ray photoevaporation driven by the central star and external FUV photoevaporation caused by irradiation form massive stars in the environment. In contrast, we here only consider the impact of photoevaporation rates driven by the central star that change in time due to stellar evolution. Our choice of calling the evolutionary pathway  outside-in just refers to the fact that, after the gap opening, it is the outer discs that dissipate faster than the inner ones.

We find outside-in disc dispersal for discs around massive stars of $2.5$ and $3\Msun$ and assuming intermediate or low viscosity parameters.
For these stars the gap is opened once the star approaches the main sequence (at ages of $\sim 1-2$\,Myr) and FUV photoevaporation becomes very efficient due to high effective temperatures. 
The X-ray photoevaporation is not strong enough here to open the gap before it drops, due to the contraction of the convective envelope, approximately at the same time FUV irradiation grows (see for example the S42 case in figs. \ref{Fig:Perfiles-y-Mpunto-Alpha3}). Instead, for low-mass discs, the gap still opens as usual due to X-ray photoevaporation. 
In short, outside-in disc clearance only happens for intermediate or low viscosities as otherwise
the viscous timescales of the inner discs becomes too short and the inner disc can not outlive the outer one. 

The transition from homogeneous to outside-in evolution is very clear for high and intermediate mass discs, but less so for the least massive discs 
(see figs.\,\ref{Fig:Perfiles-y-Mpunto-Alpha3} and \ref{Fig:Perfiles-y-Mpunto-Alpha4}).
Moreover, this transition seems to shift towards more massive stars as the mass of the disc decreases. In S42, the outside-in evolution is very evident while the evolution of S40 resembles none of the so far describe evolutionary pathways. 
We describe the corresponding evolution in detail in the following section. 

\subsection{A transition phase between homogeneous and outside-in evolution: revenant discs}
\label{Sec:RebornDisks}

While the evolutionary pathways we described so far are relatively easy to understand and depend mostly on the difference of the viscous time scale of the inner disc and the photoevaporation time scale of the outer disc, we observe a more complex but interesting evolution that occurs for initial parameters in between those that lead to homogeneous evolution and those that cause outside-in disc clearance.  
 
In these cases, X-ray photoevaporation opens a gap which leads to the formation of an inner and an outer disc, as in most of the evolutionary scenarios discussed before. However, for the parameters corresponding to our models S27 (fig. \ref{Fig:Perfiles-y-Mpunto-Alpha4}), S33 and S40 (fig. \ref{Fig:Perfiles-y-Mpunto-Alpha3}) highlighted with red simulation numbers, the outer and inner disc evolve independently for some time after the gap opening but manage to close the gap and form a continuous disc again.
 
As a fully connected disc structure is re-established in these evolutionary scenarios we call them 'revenant' disc evolution. 

\begin{figure*}
   \centering
   \includegraphics[angle=-90,width=1.\textwidth]{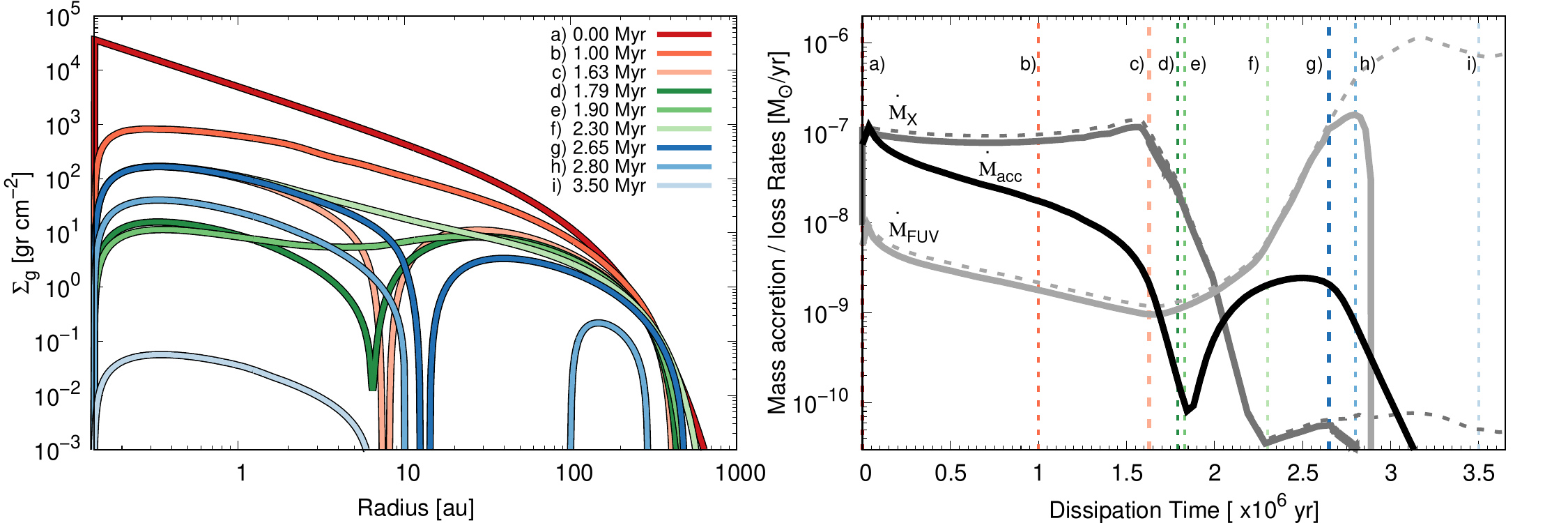}
   \caption{Time evolution of the gas surface density (left panel) and of the mass accretion and photoevaporation mass-loss rates (right panel) for the S33 simulation that we classify as a revenant disc. The redish, greenish and blueish profiles describe Stages I, II and III, respectively, for the evolution of this kind of disc (see Sec. \ref{Sec:RebornDisks}). The vertical colored and dotted lines in the right panel denote the same times of the gas profiles of the left figure. Particularly, the thicker lines denoted by 'b' and by 'g' show the positions of the first and second gap openings. It is important to note here that $\dot{M}_{\text{acc}}$ is computed at the inner border of the disc, and that $\dot{M}_{\text{X}}$ and $\dot{M}_{\text{FUV}}$ are integrated along the disc. Thus, at the time of gap openings (vertical lines c and g), these curves are not representing the situation at that specific gap opening locations.
   As in figs. \ref{Fig:Perfiles-y-Mpunto-Alpha2},\ref{Fig:Perfiles-y-Mpunto-Alpha3} and \ref{Fig:Perfiles-y-Mpunto-Alpha4}, the dashed curves correspond to the mass loss rates integrated along the disc domain while the solid ones (gray and light gray) are computed considering the available mass in the disc. A movie showing the time evolution of this disc can be downloaded from \protect\url{https://github.com/paularonco/Movies}.}
    \label{fig:RebornDisk}
   \end{figure*}

This particular evolution is the result of an interplay between the time dependence of X-ray and FUV photoevaporation and viscous evolution.  
In what follows we describe the physics of this new type of evolution, which is a direct consequence of considering the effects of stellar evolution, by separating the evolution 
into three different phases. 
We refer in particular to model S33 (the disc of $0.25\Msun$ with $\alpha$-viscosity parameter of $0.0025$ around a star of $2.5\Msun$ in fig. \ref{fig:RebornDisk}). 
In fig. \ref{fig:RebornDisk} we present the time evolution of the gas surface density profiles in the left panel and the global photoevaporation mass-loss rates $\dot{M}_{\text{X}}$ and $\dot{M}_{\text{FUV}}$ as well as the mass accretion rate $\dot{M}_{\text{acc}}$ in the right panel. 

\textbf{Phase I:} From the beginning the disc evolves as usual due to viscous accretion and mainly X-ray photoevaporation. Within the first Myr of evolution, labelled a and b in the left panel of fig. \ref{fig:RebornDisk}, $\dot{M}_{\text{acc}}$ decreases with time, 
while the X-ray photoevaporation rate, which for a $2.5\Msun$ star is effective beyond 6.25 au (see eq. \ref{eq:XRrate}), even slightly increases which leads to the typical formation of a gap in the disc at $\sim 8$ au and at 1.63 Myr (evolution from profiles b to c in fig.\ref{fig:RebornDisk}). 
During phase I, X-ray photoevaporation is always the dominant mechanism in the disc mass removal. 

\textbf{Phase II:} 
Shortly after the gap opening, $\dot{M}_{\text{X}}$ decreases sharply while $\dot{M}_{\text{FUV}}$ remains low. This implies that the total photoevaporation rate dramatically decreases. In other words, the situation that led to the gap opening, that is, photoevaporation locally dominates over viscous mass transport at the gap opening location, is no longer present. Therefore, instead of leading to a separated dispersal of the inner and the outer disc, viscous evolution of both the inner and the outer disc lead both parts to reconnect and to fill the gap (see evolution from profiles c 
to d in the left panel of fig. \ref{fig:RebornDisk}). We note here that, as $\dot{M}_{\text{acc}}$ is computed at the inner border of the disc, and as $\dot{M}_{\text{X}}$ and $\dot{M}_{\text{FUV}}$ are integrated along the disc and not computed at a certain position, the right panel of fig. \ref{fig:RebornDisk} is not representing the situation at a specific position (the gap opening locations for example) but rather what happens globally.
For model S33, both the inner and outer discs 
are separated for $\tau_{\text{sep}}\sim$ 0.16 Myr and the maximum width of the gap reaches $a_{\text{max}}=1.64$~au.

During the subsequent evolution (from profiles d to e and f in fig. \ref{fig:RebornDisk}) the local dominance of viscous processes around the gap location leads to a surface density profile corresponding to that of a stationary disc (constant mass flow throughout the disc). While reestablishing this configuration, 
the accretion rate onto the central star $\dot{M}_{\text{acc}}$ varies substantially. First, it reaches a minimum value (shortly after profile e in fig.\ref{fig:RebornDisk}) but then rapidly increases until the disc reestablishes a quasi-stationary stage and the accretion rate stays nearly constant (between profiles f and g). 

\textbf{Phase III:} The 'revenant' disc (established at time f in fig.\ref{fig:RebornDisk}) 
continues its evolution 
driven by viscous accretion and photoevaporation. 
At this stage, between the profiles f and i, $\dot{M}_{\text{X}}$ is negligible 
but $\dot{M}_{\text{FUV}}$ increases significantly. 
As for a $2.5\Msun$ star FUV photoevaporation is efficient only beyond 10 au (eq. \ref{eq:FUVrate}), a new gap is opened now at $\sim\,13$\,au at an age of $2.65$\,Myr. The disc again splits into an inner and an outer component and subsequently dissipates through outside-in evolution in $\sim$3.65 Myr.

\begin{table*}
\caption{In this table we show some characteristics of the special revenant discs describe in sec. \ref{Sec:RebornDisks}. $R_{\text{X}}$ and $\tau_{\text{X}}$ represent the location, in au, and the time, in Myr, of the gap opened by X-ray photoevaporation. $a_{\text{max}}$ represents the maximum distance of separation between the inner and outer discs formed after the gap opens, and $\tau_{\text{sep}}$ the time both discs remain apart before joining together again. $R_{\text{Reb}}$ and $\tau_{\text{Reb}}$ represent the location and time at were the disc is rebuilt. $R_{\text{FUV}}$ and $\tau_{\text{FUV}}$ denote the location and time of the second gap opened by FUV photoevaporation, and $\dot{M}_{\text{acc}}$ variability means if there is a significant change in the accretion rate.} 
\begin{center}
  \begin{tabular}{|c|c|c|c|c|c|c|c|c|c|}
    \hline
       Sim & $R_{\text{X}}$ [au] & $\tau_{\text{X}}$ [Myr] & $a_{\text{max}}$ [au] & $\tau_{\text{sep}}$ [Myr] & $R_{\text{Reb}}$ [au] & $\tau_{\text{Reb}}$ [Myr] & $R_{\text{FUV}}$ [au]& $\tau_{\text{FUV}}$ [Myr] &  $\dot{M}_{\text{acc}}$ variability\\
      \hline
       S27 & 7.19 & 2.96 & 10.96 & 1.50 & 8.25 & 4.46 & 8.43 & 4.88 & No \\      
       S33 & 7.77 & 1.63 & 1.64 & 0.16 & 6.5 & 1.79 & 13.47 & 2.66 & Yes \\
       S40 & 10.74 & 0.41 & 5.97 & 0.19 & 8.75 & 0.60 & 14.5 & 1.02 & Yes \\
       \hline
       \hline
       S35* & 9.63 & 1.23 & 16.41 & x & x & x & $\sim13.5^*$ & $\sim2.20^*$ & No \\
       S38* & x & x & x & x & x & x & 15.43 & 1.03 & Yes \\
      \hline
   \end{tabular}
  \end{center}
  \label{Tab2}
\end{table*}

We define as revenant disc evolution those models that predict an initial gap opening through X-ray photoevaporation, a full recovery of the disc  
when X-ray photoevaporation drastically drops, and a second gap opening due to FUV photoevaporation and subsequent outside-in disc dispersal. 
As revenant disc evolution represents a completely new evolutionary pathway for the gas component of protoplanetary discs, we present in table \ref{Tab2} characteristic values of the revenant disc models we identified, that is model 
S27, S33, and S40. 
The time a given disc is divided into inner and outer disc due to X-ray photoevaporation 
($\tau_{\text{sep}}$) as well as the maximum width of the predicted gap 
$a_{\text{max}}$ strongly depend on the initial disc parameters.  

In addition, the variation of the mass accretion rate predicted by model S33 (see detailed description above) is not
a defining feature of revenant disc evolution. 
In model S27 the inner and outer discs reconnect after the first gap opened but the disc has not enough time to re-establish a quasi-stationary state (which requires to establish the corresponding radial dependence of the surface density) 
since FUV photoevaporation quickly becomes efficient enough to open the second gap. 

While we define revenant disc evolution as the sequence of X-ray gap opening, filling this gap, and finally classical outside-in disc dispersal due to FUV photoevaporation, the processes driving this evolution also affect discs that do not strictly follow this sequence.  Examples for this are our simulations S35* and S38* (highlighted with red simulation numbers S35* and S38* in figs. \ref{Fig:Perfiles-y-Mpunto-Alpha4} and  \ref{Fig:Perfiles-y-Mpunto-Alpha2}, respectively) whose main characteristics are also listed in table \ref{Tab2}. 

Model S35* describes disc evolution during which the inner and outer disc almost manage to reconnect but finally do not. After the first gap is opened, the inner and outer discs achieve a maximum separation of $\sim$16 au. Then, as $\dot{M}_{\text{X}}$ sharply decrease, both disc components start to approach each other reaching a minimum separation of $\sim$1.70\,au at about 2.25\,Myr. Before the gap can be fully closed, however, $\dot{M}_{\text{FUV}}$ increases steeply and, as a consequence, FUV photoevaporation does not open a new gap but keeps open the first one until both discs completely dissipate due to viscous accretion and FUV photoevaporation. 

The disc evolution described by model S38* predicts variations of the accretion rate ($\dot{M}_{\text{acc}}$) similar to that described above for the revenant disc model S33 (see fig. \ref{fig:RebornDisk}).  
This occurs because the gas surface density decays in the region where $\dot{M}_{\text{X}}$ is most effective (around 8\,au). This reduces the mass flow to the inner parts of the disc and thus also the accretion rate ($\dot{M}_{\text{acc}}$). 
However, the decrease in $\dot{M}_{\text{X}}$ sets in before the inner and outer disc are fully separated and consequently the discs rearranges the gas surface density and the accretion rate increases until FUV photoevaporation becomes the main dissipation mechanism and the disc dissipates. 

\subsection{Dissipation Timescales}
\label{sec:Dissipation_timescales}
As the formation of gas giant planets requires the presence of a gas rich protoplanetary disc, 
an important quantity that relates disc evolution and gas giant planet formation is the disc dissipation timescale. 
In the context of dissipation time scales it is important to note that our simulations stop either if the disc mass is lower than $10^{-8}\Msun$ or if the disc lifetime is higher than 10 Myr. For this exercise and to compare with \citetalias{Kunitomo2021}, we extended our grid to discs around $4\Msun$ stars. All discs around stars more massive than $3$\,\Msun~dissipate in outside-in fashion. 

\begin{figure}
   \centering
   \includegraphics[angle=-90,width=0.99\columnwidth]{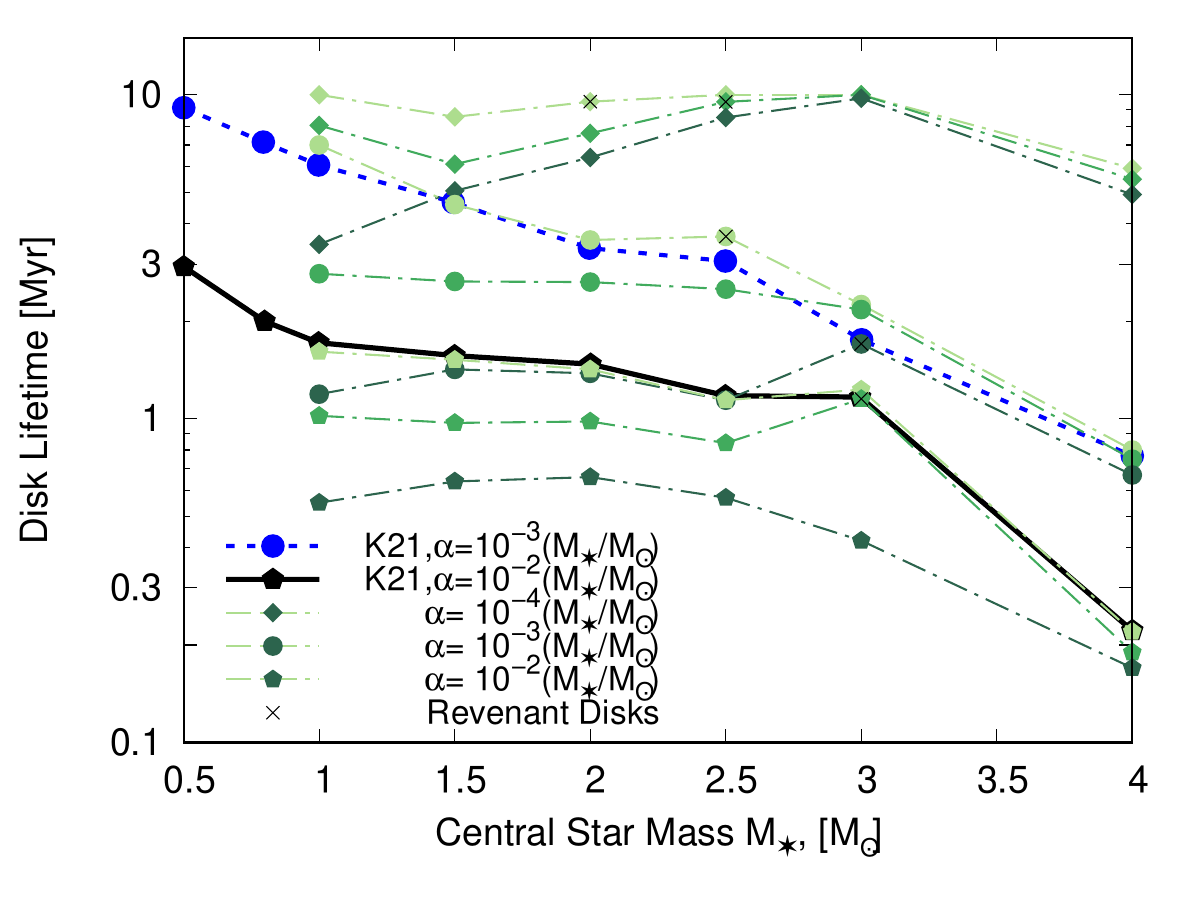}
   \caption{Dissipation timescales as a function of the central star mass for discs with $\alpha_0=10^{-2}$ (pentagons), $\alpha_0=10^{-3}$ (circles) and $\alpha_0=10^{-4}$ (diamonds), for low (dark green), intermediate (green) and high-mass (light-green) discs. Those discs marked with a black cross are denoting revenant discs. The blue and black curves are representing the results of \citet{Kunitomo2021} for discs with $\alpha_0=10^{-2}$ (pentagons) and $\alpha_0=10^{-3}$ (circles), respectively (see also their fig. 13).
   The general trend is that the disc lifetimes decrease with increasing stellar mass with the two exceptions being low-viscosity discs and revenant discs.}
    \label{fig:Tiempos}
   \end{figure}

Figure\,\ref{fig:Tiempos} shows the disc dissipation timescales as a function of stellar mass for all our simulations and provides clear evidence that for discs with the same viscosity 
the dissipation timescales decrease with the disc mass. Figure\,\ref{fig:Tiempos} also shows the dissipation timescales found by \citetalias{Kunitomo2021}, for their discs with $\alpha_0=10^{-2}$ and $\alpha_0=10^{-3}$. The disc life times we obtain for massive discs and $\alpha_0=10^{-2}$ and $\alpha_0=10^{-3}$ are in good agreement with those 
of \citetalias{Kunitomo2021} despite the different evolutionary pathways that we identified.

In general, we find a trend for the disc dissipation timescales that is in agreement with previous estimates and observations, that is, our model predicts a decrease of the disc life times with increasing stellar mass. A similar trend has been previously predicted by early modeling efforts \citep{KennedyKenyon2009} as well as confirmed by observational surveys based on near and mid infrared data \citep{Yasui2014,Ribas15}. 

In contrast to the general trend of decreasing dissipation time scales for increasing stellar mass, 
we find exceptions in two cases. 
First, assuming small values for the viscosity parameter for stars in the mass range $1.5\Msun< \Mstar \le 3\Msun$. In this region of the parameter space the dissipation timescale increases towards higher stellar masses. This happens because in these cases, the inner discs formed after the gap opening, which are more massive and larger than the ones around lower mass stars, are harder to remove due to the low viscous evolution timescales. If the stellar mass is larger than $3$\Msun~all the discs follow the general trend of shorter dissipation time scales. This is a consequence of more massive stars having much higher irradiation temperatures and thus causing lower viscous timescales.  
Despite the fact that these long-lived discs in the mass range $1.5-3$\Msun~defy the general rule, they may exist in nature. Some long-lived discs have been found around A type stars \citep[see][and references there in]{Stapper2022} and some of the ring structures in the DSHARP discs are better reproduced with low-viscosity values \citep{Dullemond2018}.

Second, when the evolutionary pathway switches from continuous to revenant disc dissipation (the models corresponding to the latter are marked with a black crosses in fig.\,\ref{fig:Tiempos}),
discs dissipate slower. This feature appears independent of the assumed viscosity parameter. 
The reason behind the observed increase in the disc life time is that between the time of the first and the second gap opening (this is between profiles c and h in the left panel of fig. \ref{fig:RebornDisk}), both the X-ray and FUV photoevaporation rates are low, as is the viscous accretion rate. As a consequence, during this period of time the mass of the disc remains almost constant. This effect, which is a direct consequence of considering stellar evolution in our model, causes these revenant discs (between homogeneous and outside-in discs) to last longer than expected from the general trend. 
Despite the fact that the extra lifetime given to revenant discs is relatively small (a few hundred thousand years), it might still be relevant for the dust evolution and thus the planet formation process. We will discuss possible implications of evolutionary sequences on planet formation in more detail in the following 
section.

\label{sec:timescales}

\section{Discussion}

We have analysed the evolution of protoplanetary discs around intermediate mass stars (from $1$ to $3\,\Msun$) assuming photoevaporation rates that change in time thanks to the stellar evolution and stellar spectra calculations provided by \citetalias{Kunitomo2021} (see their Table 1).

We found that the evolutionary pathway taken by the discs changes with increasing stellar mass from inside-out (for the lowest mass stars) to outside-in disc dispersal (for the highest mass stars). 
In between these two extremes we find discs in which the outer and inner disc disappear on similar time scales (an evolution that we call homogeneous disc dispersal) and discs in which the outer and inner disc reconnect after the first gap through X-ray photoevaporation formed and which finally disappear driven by viscous accretion and FUV photoevaporation. 
At which stellar mass exactly these transitions occur depends on the assumed viscosity parameter and initial disc mass.

\begin{figure}
   \centering
   \includegraphics[angle=0,width=0.9\linewidth]{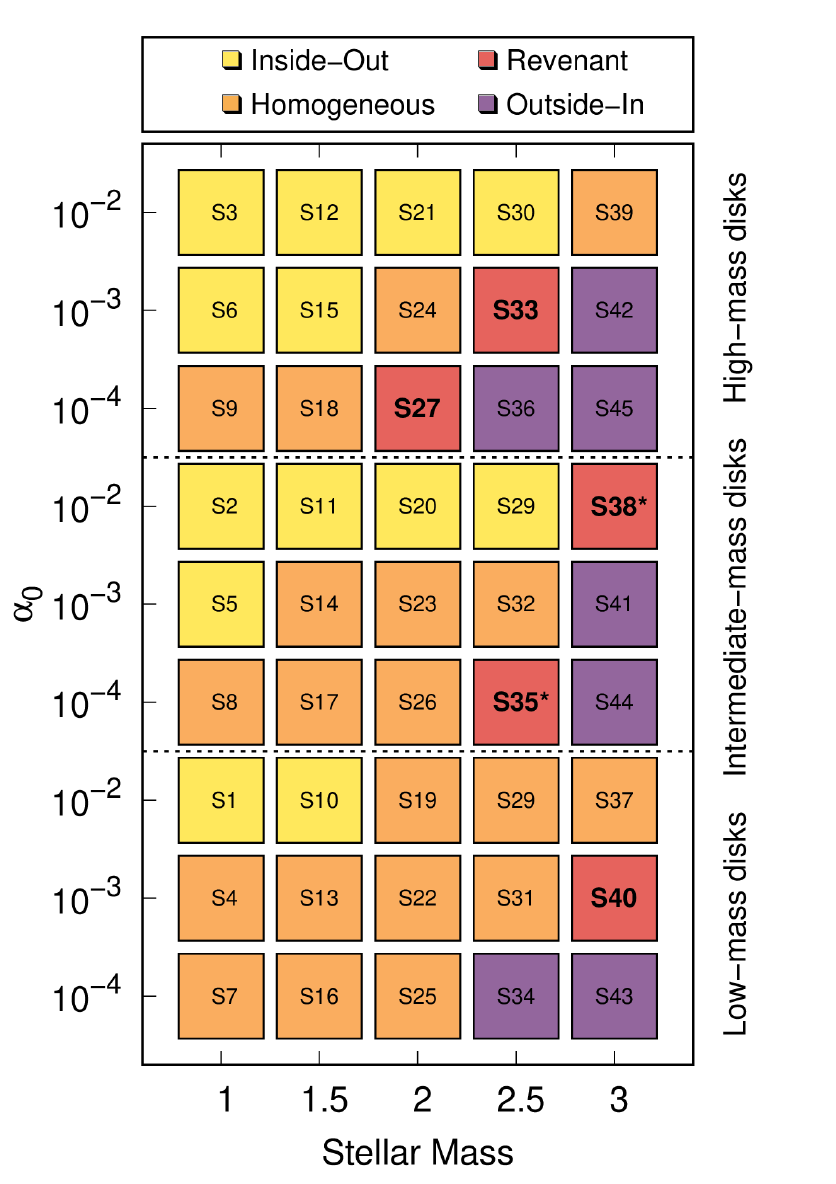}
   \caption{Summary of the findings described in sec. \ref{Sec:RebornDisks}. Each labeled square represents one of our simulations. The different colors represent the different evolutionary pathways, being yellow the classical inside-out evolution, orange the homogeneous evolution, red the revenant disc evolution (also highlighted in bold face), and violet the outside-in evolution. The red cases marked with a * are those described in sec. \ref{Sec:RebornDisks} that do not strictly follow the revenant disc evolution sequence. Unlike in figs. \ref{Fig:Perfiles-y-Mpunto-Alpha2}, \ref{Fig:Perfiles-y-Mpunto-Alpha3} and \ref{Fig:Perfiles-y-Mpunto-Alpha4}, the simulations here are organized by disc mass, and each row represents the different $\alpha_0$ viscosity parameters. The transition between the different pathways can be better appreciated this way.}
    \label{Fig:Cuadros}
    \end{figure}

Figure \ref{Fig:Cuadros} summarizes our findings and clearly shows that around stars of $\lesssim1.5$\Msun~ we only find inside-out and homogeneous disc evolution, while revenant and outside-in disc evolution are expected to occur only around stars more massive than ($\gtrsim2$\Msun). Assuming a large initial disc mass and a small value for the viscosity makes both outside-in and revenant disc evolution more likely. 

To the best of our knowledge, these different evolutionary pathways for the dissipation of protoplanetary discs around intermediate mass stars have not been identified in previous works. 
In what follows we compare our results with those found previously, discuss the potential effects of considering other X-ray photoevaporation rates and examine the outcomes of planet formation around intermediate mass stars from a theoretical and observational point of view. Moreover, we explore the connection between disks and planets around 1-3\Msun~stars to polluted white dwarfs and, in conclusion, contemplate possible implications for the planet formation process.

\subsection{Comparison with previous works}

\citet{Gorti2009} were the first to incorporate time dependent FUV photoevaporation and (constant) EUV and X-ray photoevaporation into disc model evolution. 
They predicted the corresponding evolution of protoplanetary discs around different stellar masses and found classical inside-out disc dispersal in all their models and that the discs around stars more massive than $\sim3~\Msun$ are short lived ($10^5$\,yrs). 

This early work was complemented more recently by \citetalias{Kunitomo2021} who incorporated 
time dependent X-ray and FUV photoevaporation from stellar evolution. \citetalias{Kunitomo2021} found that stellar evolution plays an important role and that, depending on the assumed viscosity parameter, disc dispersal times can be longer than those predicted by \citet{Gorti2009}. 
Both, \citet{Gorti2009} and \citetalias{Kunitomo2021} only describe inside-out disc dissipation.

The general trend for shorter lifetimes of protoplanetary discs around more massive stars has recently also been confirmed by
\citet{Komaki2021}. These authors describe disc dispersal of discs around stars for stellar masses from $0.5-7$\,\Msun~and use (constant) photoevaporation rates with a different radial dependence as those used by \citetalias{Kunitomo2021}. The same trend has been confirmed in \citet{Komaki2023}, who also considered magnetohydrodynamic
(MHD) winds as a disc dispersal mechanism.

We used the same prescriptions for the X-ray and FUV photoevaporation rates as \citetalias{Kunitomo2021}, disregarding the effects of the EUV rates, but in contrast to them (and to \citet{Komaki2021,Komaki2023}) fully solved the vertical structure equations considering a more detailed irradiation temperature (see eq. \ref{eq:temp_irrad}) and investigated a finer grid of initial disc parameters.  

The dissipation time scales we obtain are almost identical to those found previously (see fig.\ref{fig:Tiempos}), that is, about an order of magnitude longer than the early estimate by \citet{Gorti2009}.  

For the lowest mass stars, with \Mstar~ $\lesssim 2$ \Msun~we find the well known inside-out disc dispersal, previously described by several authors assuming constant values of high-energy emission from the central star \citep{Alexander2006a, Gorti2009, Owen2010, Owen2012, Coleman2022} and considering stellar evolution \citepalias{Kunitomo2021}. Particularly for the $\alpha$ viscosity parameter case $10^{-2}$ (fig. 
\ref{Fig:Perfiles-y-Mpunto-Alpha2}), our simulation S3 is comparable to the one \citetalias{Kunitomo2021} present in their fig.\,15.

In addition, thanks to the finer grid of initial conditions, we identified different evolutionary 
pathways of disc dispersal for stars more massive than $\sim2$\,\Msun. For the largest stellar masses assumed in our simulations we found that the outer disc may disappear before the inner disc. 
Such an outside-in disc dispersal has previously been predicted only for external photoevaporation from the nearby high-mass stars \citep{RichlingYorke1998,HaworthClarke2019,Coleman2022}. Despite \citetalias{Kunitomo2021} were the first including stellar evolution in their disc models, they did not report this kind of evolution for stars more massive than $\sim2$\,\Msun.
Noteworthy the case presented in their fig. 8 corresponds to initial conditions very similar to those of our case S39, where we find an homogenous disc evolution. We believe these differences, as described in detail in App. \ref{AppA}, are directly related with the differences between our (eq. \ref{eq:temp_irrad}) and their irradiation temperature (see eq. 6 in \citet{Kunitomo2020}).

The possibility of different evolutionary sequences for disc dispersal around intermediate mass stars that we identified in this work, solely caused by time-dependent photoevaporation from the central star, are completely new and might have deep implications for the planet formation process. 

\subsection{Alternative X-ray Photoevaporation Rates}
\label{sec:PhotoModels}

As mentioned in sec. \ref{sec:photoevaporation_rates}, to compute the X-ray photoevaporation rates for primordial discs we followed the same procedure as \citetalias{Kunitomo2021}, who used a simplified version of the photoevaporation rate proposed by \citet{Owen2012}.

In their section 3.2 (and also in 5.4), \citetalias{Kunitomo2021} state that the photoevaporation rates proposed by \citet{Owen2012} have recently been questioned by \citet{WangGoodman2017} and \citet{Nakatani2018a}. The latter authors performed radiation-hydrodynamic simulations with ray-tracing radiative transfer and consistent thermochemistry, and suggest that the results from \citet{Owen2012} might be overestimating the X-ray photoevaporation rates. However, analytical prescriptions for the photoevaporation rates derived \citet{WangGoodman2017} and \citet{Nakatani2018a} are currently not available and it is therefore impossible to incorporate them in our disc evolution models. The polynomial fits to the results of \citet{Nakatani2018a} by \citet{Komaki2021} cover the stellar mass range we would need but are computed for fixed EUV, X-ray and FUV luminosities that do not evolve in time. Therefore they also do not serve our purpose. The absence of prescriptions for the rates predicted by \citet{WangGoodman2017} and \citet{Nakatani2018a}
that we could use in our study is unlikely to affect the validity of our results. \citet{Sellek2022} demonstrated that the main factor causing the differences in the photoevaporation rates predicted by \citet{Owen2012} and \citet{WangGoodman2017}, is the insufficient presence of photons in the softer segment of the X-ray spectrum considered by \citet{WangGoodman2017}. 
A more realistic spectrum therefore could minimize the differences between both models and the X-ray photoevaporation rates predicted by \citet{WangGoodman2017} could become as high as the ones predicted by \citet{Owen2012}.
 
Independently, \citet{Picogna2019} and \citet{Ercolano2021} improved the X-ray prescriptions derived by \citet{Owen2012}, and later \citet{Picogna2021} provided a relation that included a dependence on the stellar mass. However, their prescriptions do not provide suitable fits for stars with masses greater than $1\Msun$ and are therefore also not useful for the context of this paper. 

\begin{figure*}
   \centering
   \includegraphics[angle=-90,width=0.99\textwidth]{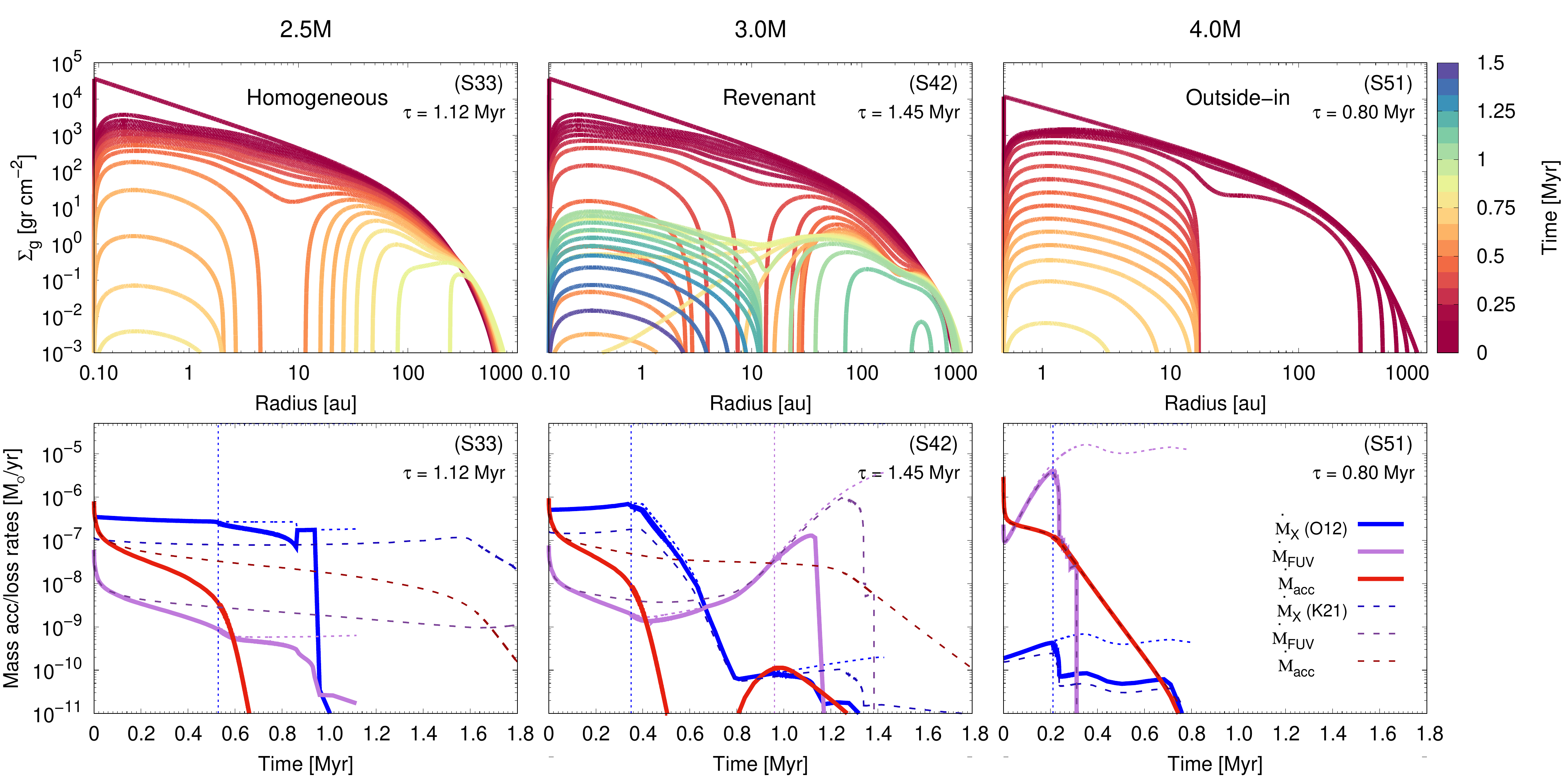}
   \caption{Time evolution of the gas surface density profiles (top panels) and of the mass accretion and photoevaporation mass-loss rates (bottom panels) for massive discs around stars with masses between $2.5\Msun$ (left panel), $3\Msun$ (middle panel) and $4.0\Msun$ (right panel) with $\alpha_0 = 10^{-3}$. In this case, the X-ray photoevaporation rate for primordial discs has been computed by directly using eq. B1, B2 and B3 from Appendix B in \citet{Owen2012}. As in figs. \ref{Fig:Perfiles-y-Mpunto-Alpha2}, \ref{Fig:Perfiles-y-Mpunto-Alpha3} and \ref{Fig:Perfiles-y-Mpunto-Alpha4}, the short dashed blue and lilac curves represent the mass loss rates but integrated along the disc domain. For comparison we also plotted the mass accretion and photoevaporation mass-loss rates (represented by the long dashed red, blue and lilac curves) computed by considering the X-ray photoevaporation rate for primordial discs as in \citetalias{Kunitomo2021}. Our general result seems to be a direct consequence of considering the effects of stellar evolution on disc evolution as they are apparently independent of the details of modeling X-ray photoevaporation. We still predict homogeneous, revenant and outside-in evolution with the only difference being that the transitions from one case to the next shifts to larger stellar masses.}
    \label{Fig:Owen}
   \end{figure*}

Thus, for the time being, the only alternative to the prescription of \citetalias{Kunitomo2021} for X-ray photoevaporation of primordial discs is to consider eq. B1, B2 and B3 from \citet{Owen2012}. To evaluate the robustness of the evolutionary pathways that we identified, we computed the time evolution of the gas surface density for massive and intermediate-viscous discs (i.e., assuming $\text{M}_\text{d}=0.1\Msun(\Mstar/\Msun)$ and $\alpha_0=10^{-3}$) around stars of 
$2.5-4$\Msun~adopting the X-ray photoevaporation rates derived by \citet[][their eq. B1, B2 and B3]{Owen2012}. Figure \ref{Fig:Owen} shows the time evolution of the gas disc profiles (top) and the time evolution of the mass accretion and mass loss rates due to photoevaporation for simulations S33, S42 and S51 (solid lines in the bottom panels). The initial conditions considered for S51, around a 4$\Msun$ star are: $\Rstar = $7715~$\Rsun$, $\Teff=6912~$K, $R_{\text{int}}=0.5~$au, $R_{\text{c}} = 95.13~$au, $\alpha = 4\times 10^{-3}$ and $\text{M}_{\text{d}}=0.4~\Msun$. 

The overall behaviour remains identical, that is, we identify homogeneous, revenant and outside-in disc dispersal. It therefore seems that discs around intermediate mass stars follow different evolutionary pathways as a consequence of changes in the photoevaporation rates as the host star evolves. 

However, as expected, the results are not identical. In what follows we describe the  differences.
First, the transition from one evolutionary sequence to the next one occurs at larger stellar masses (as compared with what happens in fig. \ref{Fig:Perfiles-y-Mpunto-Alpha3}). 
Second, the dissipation timescales for stars of 2.5\Mstar~and 3\Mstar are much shorter. This could be related to \citet{Owen2012} possibly overestimating the photoevaporation rates. 
These disc lifetimes could be extended if, instead of calculating the mass loss rate following eq. B1 in \citet{Owen2012}, one would consider eq. 9 in \citet{Ercolano2021}, where they show that for high X-ray luminosities ($L_{\text{X}}\sim10^{31}$), a plateau is reached (see their fig. 5).
For comparison we also plotted in fig.\,\ref{Fig:Owen}, as long dashed lines, the time evolution of the mass accretion and photoevaporation mass loss rates computed with eq. \ref{eq:XRrate} (see also fig. \ref{Fig:Perfiles-y-Mpunto-Alpha3}). 
The dissipation timescales for the disc around a 4\,\Mstar star considering both cases are almost identical, and this has to do with the fact that X-ray photoevaporation is very low around these massive stars and does not play a significant role.

Finally, the revenant disc evolution obtained with the prescription from \citet{Owen2012}
is a bit different to the ones previously found since the inner disc (after the first gap opening) dissipates before it can rejoin the outer one. However, when the X-ray photoevaporation rate decays, the outer disc expands viscously inwards until it reaches the inner boundary, again forming a complete disc, which then evolves mainly by viscous accretion and FUV photoevaporation, with a second gap opened by this mechanism. As in the discs S33 (see fig. \ref{fig:RebornDisk}) and S40, this disc (S42) presents variability in the accretion rate. Also, as generally described in sec. \ref{sec:Dissipation_timescales}, this disc lasts longer than expected.  

As the last issue related to X-ray photoevaporation rates that appear in the literature, we briefly discuss the potential impact of the time and location of gap opening, and of different X-ray spectral energy distributions. 
\citet{Monsch2021} showed that considering improved X-ray photoevaporation rates (derived by \citet{Picogna2019}) the gap opens earlier and also farther from the central star than with \citet{Owen2012} prescriptions (see their fig. 1). We believe that this effect would contribute to finding both, homogeneous and revenant scenarios. Particularly, gaps opening earlier could favor the formation of revenant discs, but perhaps more likely to the one described in fig.\ref{Fig:Owen}, where there would be enough time for the inner disc to dissipate completely before the outer one expands towards the inner zone. On the other hand, \citet{Ercolano2021} showed that X-ray photoevaporation rates can be different for different X-ray luminosity spectra. 
However, the impact on the results presented in the present paper are likely negligible as noted in 
\citet[][see, in particular, their fig.\,6]{Ercolano2021}. Neither the evolution of the gas surface density nor the gap location, differ significantly from those found using the X-ray photoevaporation rates derived by \citet{Picogna2019}. 
We therefore suspect that both homogeneous and revenant disc evolution occur independent of the 
assumed spectral energy distribution of the irradiating X-rays. 
The disc viscosities and masses for which the transitions are found might, however, be affected by both, the location of the gap opening and by the X-ray spectrum. 
If, for whatever reason, the first gap opened by X-ray photoevaporation closer to the star, we still expect to find revenant disc evolution, but again, probably similar to the one we show in fig. \ref{Fig:Owen}.

\subsection{Discs and giant planets around
intermediate mass stars: observations and theory}

Observations indicate that more massive stars host more massive planets \citep{Cumming2008}, which is in agreement with theoretical predictions \citep{Kennedy2008a,Kennedy2008b}. 
Also, the frequency of giant planets increases as a function of stellar mass  
\citep{Johnson2010,Reffert2015}.

According to the most recent results, the occurrence of giant planets around intermediate mass stars peaks at $\sim1.7$\Msun~and steeply decreases towards larger masses \citep{Wolthoff2022}. However, for stellar masses larger than $\sim2.5$\Msun~low number statistics still affect the reliability of the mentioned result. 

Statistics of giant planets around intermediate mass stars must be related to the evolution of protoplanetary discs around these stars. A recent survey of Herbig stars ($1-3$\Msun), provided evidence for on average more massive and larger discs that also exist for longer than those around T\,Tauri stars \citep[see][their figures 3, 5 and 6]{Stapper2022}. In addition, discs around more massive stars show more frequently sub-structures in continuum emission in the form of rings and gaps \citep{vanderMarel2021} that might be related to the formation of giant planets.  

A potentially too simple but intriguing scenario explaining the increase of the occurrence of gas giant planets up to $1.5-2$\Msun~might be that more massive discs produce gas giants more easily at greater distances. The forming gas giant planets in these more massive discs could create pressure bumps which could prevent the dust particles to spiral-in through radial drift \citep[e.g.][]{Birnstiel12}
thereby influencing the overall disc evolution \citep{Cieza2021}.  

A reason for the observed strong decrease in the planet occurrence rate around stars more massive than $\sim2-3$\Msun~ could be that the disc lifetime around these stars is shorter than that of low-mass stars \citep{Ribas15,Luhman2022} which could suggest that perhaps discs disappear on time scales short enough to prohibit gas giant planet formation.  

Indeed, early theoretical works suggested that a combination between two effects could be the reason for the apparent absence of planets around stars more massive than $2-3$\Msun. First, the snow line is located further out, where the growth timescales for planetary formation by planetesimal accretion are longer than the migration timescales. Second, photoevaporation is more efficient as more massive stars emit larger fluxes of high energy photons. The combination of these effects may prevent planetary formation \citep{IdaLin2008,Kennedy2008a,Alibert2011}. 

However, the above would be true only for planet formation models based on planetesimal accretion which need longer disc lifetimes to allow planets to grow and to migrate inwards. 
In contrast, models based on pebble accretion predict efficient planetary growth which leads to a fast inward migration \citep{Venturini2020SE,Venturini2020Letter, Guilera2021,Drazkowska2021}. Nevertheless, \citet{Pinilla2022} recently suggested that a dramatic increase of the radial drift velocity as soon as the stellar mass exceeds $2-3$\Msun~might in fact hinder the formation of massive cores, precursors of gas giant planets even if pebble accretion is considered (see sec. \ref{Cons-planet-formation}).

In any case, the large number of gas giant planets at several au from the host star observer around $1-2$\Msun~as well as the decrease of the planet occurrence rate at stellar masses of $2-3$\Msun~ could be potentially related to the different evolutionary pathways that we discovered in this work.   

\subsection{Discs and giant planets around
intermediate mass stars: the link with polluted white dwarfs}

As previously mentioned, a reason for studying disc evolution and planet formation around intermediate mass stars stems from their most common stellar remnants: white dwarfs. 

Only a few planets and planet candidates have been detected around white dwarfs \citep{Gaensicke2019,Vandenburg2020,Blackman2021}, and just one of them is located at a separation where it could have survived the evolution of its host star \citep{Blackman2021}. The others must have either scattered inward \citep{Maldonado2021} which would imply the existence of additional planets, reached large eccentricities through angular momentum exchange with a distant tertiary \citep{Oconnor2021,Munoz2021,Stephan2021}, survived common envelope evolution \citep{Lagos2021,Lagos2023}, or are second generation planets formed in a latter circumstellar disc \citep{2010arXiv1001.0581P, 2022A&A...658A..36K,2023A&A...675A.184L}.

While the sample of confirmed planets around white dwarfs is thus currently very small, about one third of all white dwarfs show metal absorption lines \citep[e.g.][]{Koester2014} which are thought to be caused by the accretion of planetary debris produced by tidal disruption of asteroids, planetesimals, or planets \citep[e.g.][]{Jura2003,Gaensicke2006,Veras2014, Vandenburg2015,Manser2019,Malamud2020,Veras2021}. The objects producing the debris survived the evolution of the host star into a white dwarf and therefore must originate from distances of at least a few au \citep[e.g.][]{Mustill2012,Ronco2020}.

About 40\% of field WDs have masses between 0.6\Msun~ and 0.75\Msun \citep{Tremblay2016} which correspond to progenitor stars with initial masses between 1.5\Msun~ and 3\Msun\ \citep{Cummings2018,2020NatAs...4.1102M}, the mass range studied in this work.
The large number of metal polluted white dwarfs with masses below $0.7$\,\Msun~\citep[][their figure 1]{Koester2014} might imply that planet formation around stars in the mass range $1-3$\,\Msun~is potentially rather efficient. In fact, recently \cite{2021MNRAS.504.2707G} identified two WD systems with stellar masses $\sim 0.63$\Msun~ and\ 0.64\Msun~ with very bright gaseous debris discs.

The discoveries of polluted white dwarfs, planets around white dwarfs, and planets orbiting intermediate-mass stars could be potentially related to the different evolutionary pathways that we discovered in this work, and provide added impetus to further enhance our understanding of both disc evolution and the mechanisms involved in planet formation across these stars.

\subsection{Implications for planet formation}
\label{Cons-planet-formation}
\begin{figure}
   \centering
   \includegraphics[angle=-90,width=1.\columnwidth]{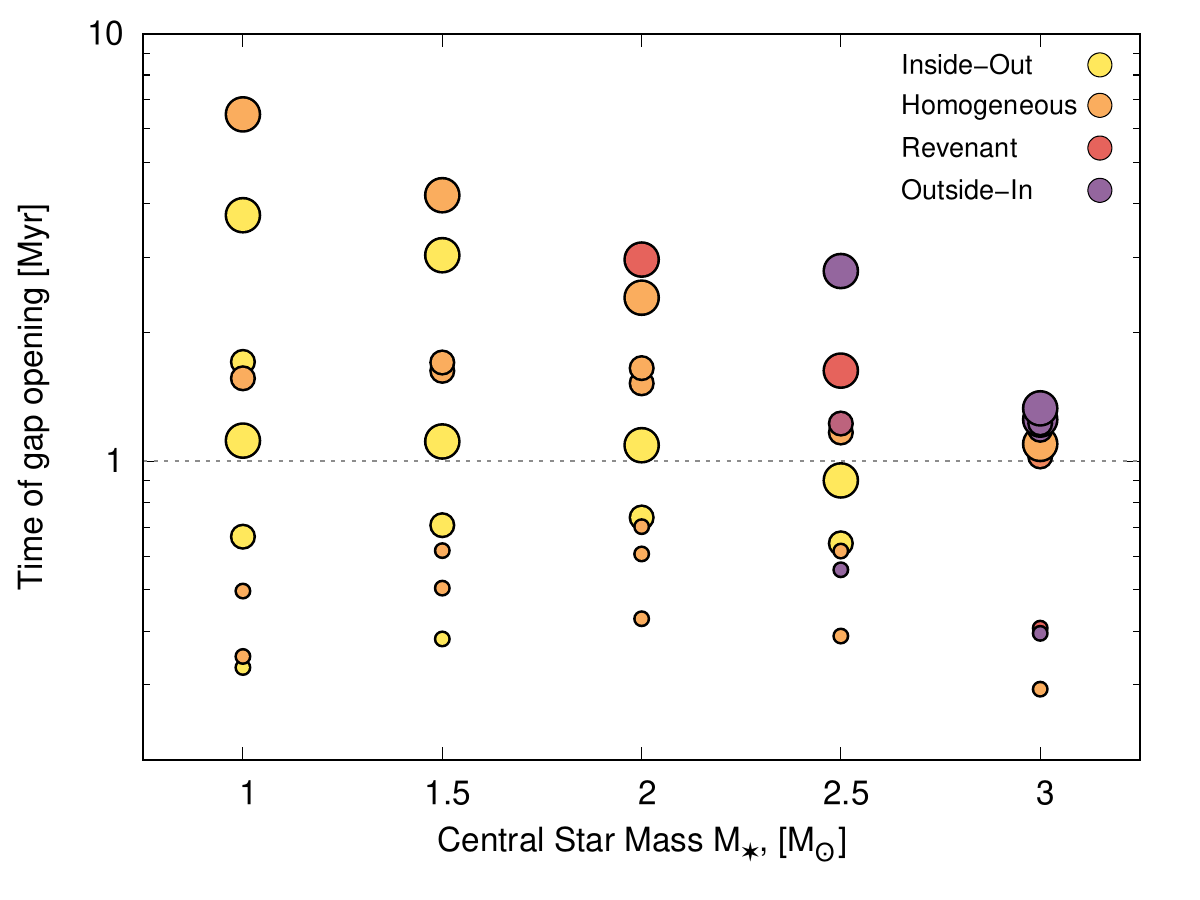}
   \caption{Time of the gap opening as a function of the central star mass for each of our simulations. As in fig. \ref{Fig:Cuadros},  the colors represent the different evolutionary pathways, being yellow the classical inside-out evolution, orange the homogeneous evolution, red the revenant disc evolution, and violet the outside-in evolution. Small dots represent low-mass discs while middle size dots and big dots represent intermediate and high-mass discs, respectively. As we can see here, it is mostly low-mass discs that open a gap in less than 1 Myr, stopping early the pebble flux from the outer parts. 
   \label{fig:Time-of-gap-opening}}
\end{figure}

When it comes to forming planets, perhaps more important than taking into account the total lifetime of the disc, is knowing if a gap is opened due to photoevaporation and particularly at what time this happens. This event will basically divide the disc in two parts and will stop the pebble flux from the outer regions towards the inner ones, likely affecting the outcomes of planet formation. Indeed, \citet{Venturini2020SE} showed that the pebble flux and the pebble surface density decline abruptly in the inner parts of the disc after the gap opening, limiting the formation of massive rocky planets ($\gtrsim5\text{M}_\oplus$). In addition, an early gap opening could also halt the migration of planets formed in the outer parts of the disc. If this happens around intermediate mass stars it could be related with the lack of close-in planets around them.

Fig.\,\ref{fig:Time-of-gap-opening} shows, for each of our simulations, the time at which the gap is opened due to photoevaporation. The colored dots refer to the type of disc pathway, as previously defined, and following fig. \ref{Fig:Cuadros}. We clearly see that low-mass discs (smaller dots) open the gap pretty early, in less than 1 Myr. However, most of the intermediate and high-mass discs (intermediate size and bigger circles, respectively), independently of their evolutionary pathway, do it later, probably allowing enough time for planets to form and to migrate inwards, especially for those formed by pebble accretion. 

\begin{figure}
   \centering
   \includegraphics[angle=-90,width=0.98\columnwidth]{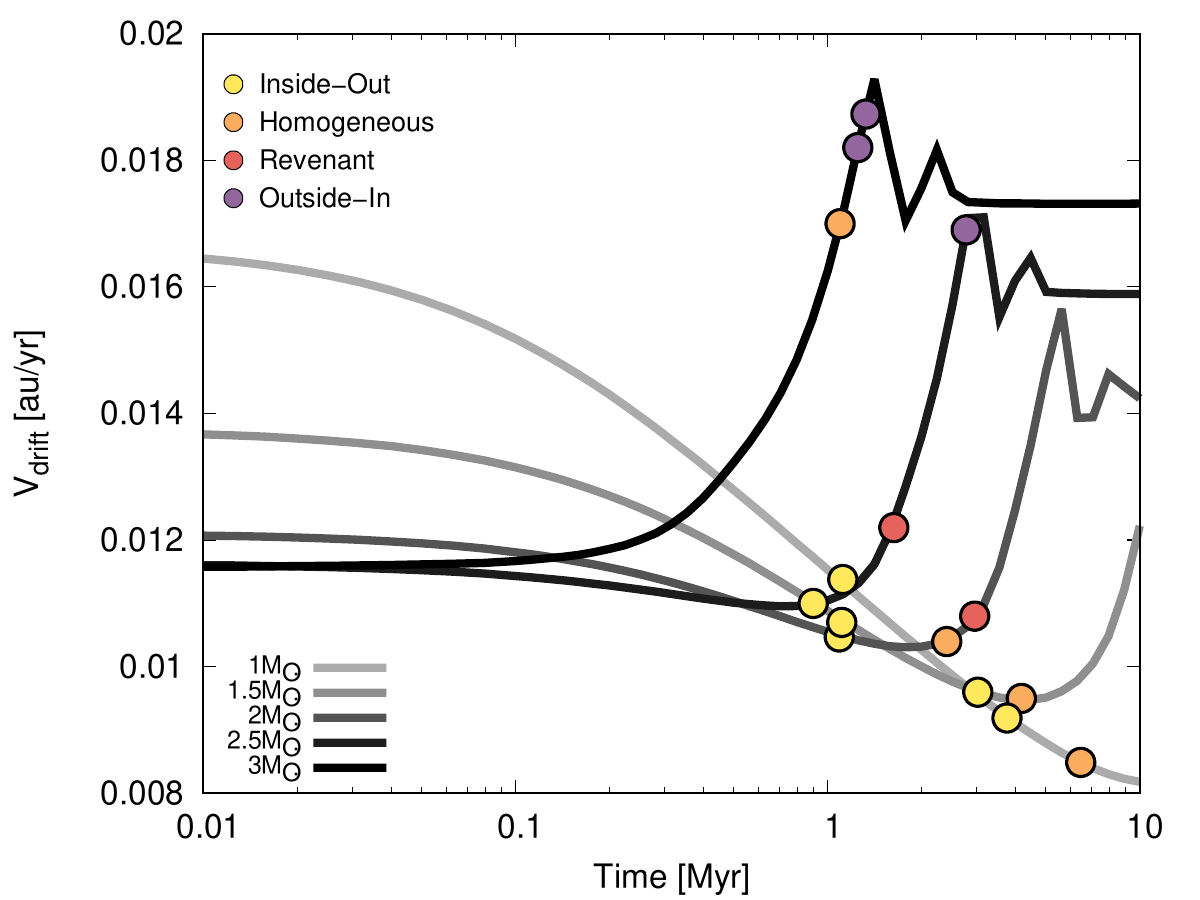}
   \caption{Time evolution of the pebble drift velocities, as a function of $\Lstar$ and $\Mstar$ \citep{Pinilla2013}, for stars with masses between 1 and $3\Msun$. The colored circles represent the time of gap opening only for those simulations with the high-mass discs. As in fig. \ref{fig:Time-of-gap-opening}, the colors represent the different evolutionary pathways.}
   \label{Fig:Vdrift}
\end{figure}

On the other hand, while \citet{Guilera2021} and \citet{Drazkowska2021} found planet formation timescales typically shorter than 1~Myr around 1\Msun~ stars, \citet{Pinilla2022} argue that planet formation around more massive stars could be limited by an increment in the pebble drift velocities due to the effects of stellar evolution. With this in mind, we show in figure \ref{Fig:Vdrift} (as in the right panel of fig. 2 in \citet{Pinilla2022}), the time evolution of the pebble drift velocities $V_{\text{drift}}$ computed as $V_{\text{drift}}\propto L^{1/4}_\star/\sqrt{\Mstar}$ \citep{Pinilla2013}, for the different stellar masses we considered using the results of Table 1 of \citetalias{Kunitomo2021}. For stellar masses between $1\Msun$ and $3\Msun$ we find good agreement with the results of \citet{Pinilla2022}. Moreover, on top of each curve we plot the time scales of gap opening of the simulations with the high-mass discs. The increment in the drift velocities affects mainly stars of $3\Msun$ in which the time of gap opening occurs close to the maximum of the $V_{\text{drift}}$ curve. This could imply that, despite the fact that in these discs the gap opens after 1 Myr (and also lives for long timescales) most of the solid component could be lost due to the pebble drift thereby limiting planet formation. For the rest of the cases (except for the S36 case, the only violet circle in the 2.5\Msun~curve), the gap opens before the significant increment in the pebble drift velocity. We draw attention to the fact that these $V_{\text{drift}}$ curves were computed assuming, as in \citet{Pinilla2022}, that all the pebbles along the disc are represented by Stokes numbers St=1. 

An important thing to highlight here is that our detailed computation of the vertical structure of the disc provides detail temperature profiles at the disc mid-plane. These temperatures are fundamental to properly compute planet migration rates in non-isothermal discs. As discussed in App. \ref{AppA}, and particularly for highly viscous discs, we find temperature values that are a factor 2 to several times lower than those of \citetalias{Kunitomo2021}. This difference could lead to different planet migration pathways along the disc.

Finally, different disc evolutionary pathways, as the ones we find among our simulations and which are a direct consequence of considering stellar evolution, could also affect the results of planet formation. 

Particularly interesting, in the context of planet formation, could be the revenant disc evolution, which could play an important role in the dust evolution and planet formation. In this case, when the first gap is opened, a pressure maximum is generated at the edge of the outer disc wherein the dust could be efficiently accumulated, triggering planetesimal and planet formation \citep[e.g.][]{Guilera2020,Jiang2023}. In addition, this pressure maximum should also act as a planet migration trap \citep[e.g][]{Masset2006, Guilera2017, Guilera2020}. Then, when $\dot{M}_{\text{X}}$ decays and the disc is rebuilt, the accumulated dust should drift inward supplying the inner part of the disc with solids and allowing those planets that might have formed in the pressure maximum to migrate inwards. 

Regarding the cases of outside-in evolution, although the gas dissipation time scales of the inner parts of these discs are pretty long, this does not necessarily imply that the same will happen with the solid component which could be strongly affected by the drift time scales and be lost long before the gas dissipates. 
We will investigate in a follow up paper in more detail the impact of the evolutionary sequences we identified here on planet formation. 

\section{Conclusions}

We incorporated stellar evolution and the corresponding time dependence of X-ray and FUV photoevaporation in our 1D+1D code {\scriptsize PLANETALP} describing the evolution of protoplanetary discs with a detailed computation of its vertical structure. We tested the impact of this effect on the evolution of discs around stars with masses covering the range $1-3$\,\Msun~and assuming different values for the initial disc mass and the viscosity parameter. 

Our main results are as follows. First, and most importantly, the evolutionary pathway of protoplanetary disc dispersal due to photoevaporation depends on the stellar mass. For low mass stars, disc evolution follows the classical inside-out clearing, but it changes to homogeneous disc evolution (that is the inner and outer disc disappear on a similar time scale), revenant disc evolution (where the inner and the outer disc are able to reconnect after the first gap opened), and outside-in disc dispersal (during which the outer disc dissipates first) as the mass of the central star increases. At which stellar mass exactly the evolutionary sequence changes depends on the assumed viscosity parameter, the assumed initial mass of the disc, and the prescription used for the X-ray photoevaporation. Second, we found that the disc dispersal time scale decreases with increasing stellar mass except for low viscosity discs and when the disc switches to revenant disc evolution. Revenant disc evolution seems to extend a bit the disc lifetime. These evolutionary pathways do not seem to depend on the considered photoevaporation prescription.

The identified changes of protoplanetary disc evolution and the changes of the discs lifetime with increasing stellar mass may have a substantial impact on our understanding of planet formation and the observed occurrence rate of giant planets around main sequence stars. Moreover, they may also have an effect on our understanding of white dwarf pollution.

We will investigate the impact of the evolutionary sequences we identified here on the dust component of protoplanetary discs (and thus on planet formation) in a subsequent publication. 

\begin{acknowledgements}
 The authors would like to thank the referee, Giovanni Picogna, for his constructive suggestions that helped to improve the manuscript. MPR, MRS and EV are grateful to the KITP Program "White Dwarfs as Probes of the Evolution of Planets, Stars, the Milky Way and the Expanding Universe" during which the development of this work began. This research was supported in part by the National Science Foundation under Grant No. NSF PHY-1748958. MPR also would like to thank Masanobu Kunitomo for sharing the data from his work to facilitate comparisons.
 MPR is partially supported by PICT-2021-I-INVI-00161 from ANPCyT, Argentina. MPR, OMG and M3B are partially supported by PIP-2971 from CONICET (Argentina) and by PICT 2020-03316 from Agencia I+D+i (Argentina). MPR, MRS, and OMG acknowledge support by ANID, -- Millennium Science Initiative Program -- NCN19\_171. EV acknowledges support from the DISCOBOLO project funded by the Spanish Ministerio de Ciencia, Innovación y Universidades under grant PID2021-127289NB-I00. MRS is further supported by Fondecyt (grant 1221059). MPR, OMG and M3B also thank Juan Ignacio Rodriguez from IALP for the computation managing resources of the Grupo de Astrofísica Planetaria de La Plata. 
\end{acknowledgements}

%
%

\bibliographystyle{aa}
\bibliography{Biblio}

\begin{thebibliography}{115}
\expandafter\ifx\csname natexlab\endcsname\relax\def\natexlab#1{#1}\fi

\bibitem[{{Alexander} {et~al.}(2006{\natexlab{a}}){Alexander}, {Clarke}, \&
  {Pringle}}]{Alexander2006a}
{Alexander}, R.~D., {Clarke}, C.~J., \& {Pringle}, J.~E. 2006{\natexlab{a}},
  \mnras, 369, 216

\bibitem[{{Alexander} {et~al.}(2006{\natexlab{b}}){Alexander}, {Clarke}, \&
  {Pringle}}]{Alexander2006b}
{Alexander}, R.~D., {Clarke}, C.~J., \& {Pringle}, J.~E. 2006{\natexlab{b}},
  \mnras, 369, 229

\bibitem[{{Alibert} {et~al.}(2011){Alibert}, {Mordasini}, \&
  {Benz}}]{Alibert2011}
{Alibert}, Y., {Mordasini}, C., \& {Benz}, W. 2011, \aap, 526, A63

\bibitem[{{Alibert} {et~al.}(2005){Alibert}, {Mordasini}, {Benz}, \&
  {Winisdoerffer}}]{Alibert05}
{Alibert}, Y., {Mordasini}, C., {Benz}, W., \& {Winisdoerffer}, C. 2005, \aap,
  434, 343

\bibitem[{{Alibert} {et~al.}(2018){Alibert}, {Venturini}, {Helled}, {Ataiee},
  {Burn}, {Senecal}, {Benz}, {Mayer}, {Mordasini}, {Quanz}, \&
  {Sch{\"o}nb{\"a}chler}}]{Alibert2018}
{Alibert}, Y., {Venturini}, J., {Helled}, R., {et~al.} 2018, Nature Astronomy,
  2, 873

\bibitem[{{Andrews} {et~al.}(2010){Andrews}, {Wilner}, {Hughes}, {Qi}, \&
  {Dullemond}}]{Andrews2010}
{Andrews}, S.~M., {Wilner}, D.~J., {Hughes}, A.~M., {Qi}, C., \& {Dullemond},
  C.~P. 2010, \apj, 723, 1241

\bibitem[{{Birnstiel} {et~al.}(2012){Birnstiel}, {Klahr}, \&
  {Ercolano}}]{Birnstiel12}
{Birnstiel}, T., {Klahr}, H., \& {Ercolano}, B. 2012, \aap, 539, A148

\bibitem[{{Blackman} {et~al.}(2021){Blackman}, {Beaulieu}, {Bennett},
  {Danielski}, {Alard}, {Cole}, {Vandorou}, {Ranc}, {Terry}, {Bhattacharya},
  {Bond}, {Bachelet}, {Veras}, {Koshimoto}, {Batista}, \&
  {Marquette}}]{Blackman2021}
{Blackman}, J.~W., {Beaulieu}, J.~P., {Bennett}, D.~P., {et~al.} 2021, \nat,
  598, 272

\bibitem[{{Burn} {et~al.}(2021){Burn}, {Schlecker}, {Mordasini}, {Emsenhuber},
  {Alibert}, {Henning}, {Klahr}, \& {Benz}}]{Burn2021}
{Burn}, R., {Schlecker}, M., {Mordasini}, C., {et~al.} 2021, \aap, 656, A72

\bibitem[{{Chiang} \& {Goldreich}(1997)}]{Chiang1997}
{Chiang}, E.~I. \& {Goldreich}, P. 1997, \apj, 490, 368

\bibitem[{{Cieza} {et~al.}(2021){Cieza}, {Gonz{\'a}lez-Ruilova}, {Hales},
  {Pinilla}, {Ru{\'\i}z-Rodr{\'\i}guez}, {Zurlo}, {Casassus}, {P{\'e}rez},
  {C{\'a}novas}, {Arce-Tord}, {Flock}, {Kurtovic}, {Marino}, {Nogueira},
  {Perez}, {Price}, {Principe}, \& {Williams}}]{Cieza2021}
{Cieza}, L.~A., {Gonz{\'a}lez-Ruilova}, C., {Hales}, A.~S., {et~al.} 2021,
  \mnras, 501, 2934

\bibitem[{{Clarke} {et~al.}(2001){Clarke}, {Gendrin}, \&
  {Sotomayor}}]{Clarke2001}
{Clarke}, C.~J., {Gendrin}, A., \& {Sotomayor}, M. 2001, \mnras, 328, 485

\bibitem[{{Coleman} \& {Haworth}(2022)}]{Coleman2022}
{Coleman}, G. A.~L. \& {Haworth}, T.~J. 2022, \mnras, 514, 2315

\bibitem[{{Cumming} {et~al.}(2008){Cumming}, {Butler}, {Marcy}, {Vogt},
  {Wright}, \& {Fischer}}]{Cumming2008}
{Cumming}, A., {Butler}, R.~P., {Marcy}, G.~W., {et~al.} 2008, \pasp, 120, 531

\bibitem[{{Cummings} {et~al.}(2018){Cummings}, {Kalirai}, {Tremblay},
  {Ramirez-Ruiz}, \& {Choi}}]{Cummings2018}
{Cummings}, J.~D., {Kalirai}, J.~S., {Tremblay}, P.~E., {Ramirez-Ruiz}, E., \&
  {Choi}, J. 2018, \apj, 866, 21

\bibitem[{{Drazkowska} {et~al.}(2022){Drazkowska}, {Bitsch}, {Lambrechts},
  {Mulders}, {Harsono}, {Vazan}, {Liu}, {Ormel}, {Kretke}, \&
  {Morbidelli}}]{Drazkowska22}
{Drazkowska}, J., {Bitsch}, B., {Lambrechts}, M., {et~al.} 2022, arXiv
  e-prints, arXiv:2203.09759

\bibitem[{{Dr{\k{a}}{\.z}kowska} {et~al.}(2021){Dr{\k{a}}{\.z}kowska},
  {Stammler}, \& {Birnstiel}}]{Drazkowska2021}
{Dr{\k{a}}{\.z}kowska}, J., {Stammler}, S.~M., \& {Birnstiel}, T. 2021, \aap,
  647, A15

\bibitem[{{Dullemond} {et~al.}(2018){Dullemond}, {Birnstiel}, {Huang},
  {Kurtovic}, {Andrews}, {Guzm{\'a}n}, {P{\'e}rez}, {Isella}, {Zhu}, {Benisty},
  {Wilner}, {Bai}, {Carpenter}, {Zhang}, \& {Ricci}}]{Dullemond2018}
{Dullemond}, C.~P., {Birnstiel}, T., {Huang}, J., {et~al.} 2018, \apjl, 869,
  L46

\bibitem[{{Ercolano} \& {Pascucci}(2017)}]{ErcolanoPascucci2017}
{Ercolano}, B. \& {Pascucci}, I. 2017, Royal Society Open Science, 4, 170114

\bibitem[{{Ercolano} {et~al.}(2021){Ercolano}, {Picogna}, {Monsch}, {Drake}, \&
  {Preibisch}}]{Ercolano2021}
{Ercolano}, B., {Picogna}, G., {Monsch}, K., {Drake}, J.~J., \& {Preibisch}, T.
  2021, \mnras, 508, 1675

\bibitem[{{G{\"a}nsicke} {et~al.}(2006){G{\"a}nsicke}, {Marsh}, {Southworth},
  \& {Rebassa-Mansergas}}]{Gaensicke2006}
{G{\"a}nsicke}, B.~T., {Marsh}, T.~R., {Southworth}, J., \&
  {Rebassa-Mansergas}, A. 2006, Science, 314, 1908

\bibitem[{{G{\"a}nsicke} {et~al.}(2019){G{\"a}nsicke}, {Schreiber}, {Toloza},
  {Gentile Fusillo}, {Koester}, \& {Manser}}]{Gaensicke2019}
{G{\"a}nsicke}, B.~T., {Schreiber}, M.~R., {Toloza}, O., {et~al.} 2019, \nat,
  576, 61

\bibitem[{{Gentile Fusillo} {et~al.}(2021){Gentile Fusillo}, {Manser},
  {G{\"a}nsicke}, {Toloza}, {Koester}, {Dennihy}, {Brown}, {Farihi},
  {Hollands}, {Hoskin}, {Izquierdo}, {Kinnear}, {Marsh},
  {Santamar{\'\i}a-Miranda}, {Pala}, {Redfield}, {Rodr{\'\i}guez-Gil},
  {Schreiber}, {Veras}, \& {Wilson}}]{2021MNRAS.504.2707G}
{Gentile Fusillo}, N.~P., {Manser}, C.~J., {G{\"a}nsicke}, B.~T., {et~al.}
  2021, \mnras, 504, 2707

\bibitem[{{Gorti} {et~al.}(2009){Gorti}, {Dullemond}, \&
  {Hollenbach}}]{Gorti2009}
{Gorti}, U., {Dullemond}, C.~P., \& {Hollenbach}, D. 2009, \apj, 705, 1237

\bibitem[{{Guilera} {et~al.}(2010){Guilera}, {Brunini}, \&
  {Benvenuto}}]{Guilera2010}
{Guilera}, O.~M., {Brunini}, A., \& {Benvenuto}, O.~G. 2010, \aap, 521, A50

\bibitem[{{Guilera} {et~al.}(2019){Guilera}, {Cuello}, {Montesinos}, {Miller
  Bertolami}, {Ronco}, {Cuadra}, \& {Masset}}]{Guilera2019}
{Guilera}, O.~M., {Cuello}, N., {Montesinos}, M., {et~al.} 2019, \mnras, 486,
  5690

\bibitem[{{Guilera} {et~al.}(2021){Guilera}, {Miller Bertolami}, {Masset},
  {Cuadra}, {Venturini}, \& {Ronco}}]{Guilera2021}
{Guilera}, O.~M., {Miller Bertolami}, M.~M., {Masset}, F., {et~al.} 2021,
  \mnras, 507, 3638

\bibitem[{{Guilera} {et~al.}(2017){Guilera}, {Miller Bertolami}, \&
  {Ronco}}]{Guilera2017Letter}
{Guilera}, O.~M., {Miller Bertolami}, M.~M., \& {Ronco}, M.~P. 2017, \mnras,
  471, L16

\bibitem[{{Guilera} \& {S{\'a}ndor}(2017)}]{Guilera2017}
{Guilera}, O.~M. \& {S{\'a}ndor}, Z. 2017, \aap, 604, A10

\bibitem[{{Guilera} {et~al.}(2020){Guilera}, {S{\'a}ndor}, {Ronco},
  {Venturini}, \& {Miller Bertolami}}]{Guilera2020}
{Guilera}, O.~M., {S{\'a}ndor}, Z., {Ronco}, M.~P., {Venturini}, J., \& {Miller
  Bertolami}, M.~M. 2020, arXiv e-prints, arXiv:2005.10868

\bibitem[{{Haworth} \& {Clarke}(2019)}]{HaworthClarke2019}
{Haworth}, T.~J. \& {Clarke}, C.~J. 2019, \mnras, 485, 3895

\bibitem[{{Hayashi}(1981)}]{Hayashi81}
{Hayashi}, C. 1981, Progress of Theoretical Physics Supplement, 70, 35

\bibitem[{{Ida} \& {Lin}(2008)}]{IdaLin2008}
{Ida}, S. \& {Lin}, D.~N.~C. 2008, \apj, 673, 487

\bibitem[{{Ikoma} {et~al.}(2000){Ikoma}, {Nakazawa}, \& {Emori}}]{Ikoma2000}
{Ikoma}, M., {Nakazawa}, K., \& {Emori}, H. 2000, \apj, 537, 1013

\bibitem[{{Jiang} \& {Ormel}(2023)}]{Jiang2023}
{Jiang}, H. \& {Ormel}, C.~W. 2023, \mnras, 518, 3877

\bibitem[{{Johansen} \& {Lambrechts}(2017)}]{JohansenLambrechts2017}
{Johansen}, A. \& {Lambrechts}, M. 2017, Annual Review of Earth and Planetary
  Sciences, 45, 359

\bibitem[{{Johnson} {et~al.}(2010){Johnson}, {Aller}, {Howard}, \&
  {Crepp}}]{Johnson2010}
{Johnson}, J.~A., {Aller}, K.~M., {Howard}, A.~W., \& {Crepp}, J.~R. 2010,
  \pasp, 122, 905

\bibitem[{{Jura}(2003)}]{Jura2003}
{Jura}, M. 2003, \apjl, 584, L91

\bibitem[{{Kennedy} \& {Kenyon}(2008{\natexlab{a}})}]{Kennedy2008b}
{Kennedy}, G.~M. \& {Kenyon}, S.~J. 2008{\natexlab{a}}, \apj, 682, 1264

\bibitem[{{Kennedy} \& {Kenyon}(2008{\natexlab{b}})}]{Kennedy2008a}
{Kennedy}, G.~M. \& {Kenyon}, S.~J. 2008{\natexlab{b}}, \apj, 673, 502

\bibitem[{{Kennedy} \& {Kenyon}(2009)}]{KennedyKenyon2009}
{Kennedy}, G.~M. \& {Kenyon}, S.~J. 2009, \apj, 695, 1210

\bibitem[{{Kluska} {et~al.}(2022){Kluska}, {Van Winckel}, {Copp{\'e}e},
  {Oomen}, {Dsilva}, {Kamath}, {Bujarrabal}, \& {Min}}]{2022A&A...658A..36K}
{Kluska}, J., {Van Winckel}, H., {Copp{\'e}e}, Q., {et~al.} 2022, \aap, 658,
  A36

\bibitem[{{Koester} {et~al.}(2014){Koester}, {G{\"a}nsicke}, \&
  {Farihi}}]{Koester2014}
{Koester}, D., {G{\"a}nsicke}, B.~T., \& {Farihi}, J. 2014, \aap, 566, A34

\bibitem[{{Komaki} {et~al.}(2023){Komaki}, {Fukuhara}, {Suzuki}, \&
  {Yoshida}}]{Komaki2023}
{Komaki}, A., {Fukuhara}, S., {Suzuki}, T.~K., \& {Yoshida}, N. 2023, arXiv
  e-prints, arXiv:2304.13316

\bibitem[{{Komaki} {et~al.}(2021){Komaki}, {Nakatani}, \&
  {Yoshida}}]{Komaki2021}
{Komaki}, A., {Nakatani}, R., \& {Yoshida}, N. 2021, \apj, 910, 51

\bibitem[{{Kraus}(2015)}]{Kraus2015}
{Kraus}, S. 2015, \apss, 357, 97

\bibitem[{{Kunitomo} {et~al.}(2021){Kunitomo}, {Ida}, {Takeuchi}, {Pani{\'c}},
  {Miley}, \& {Suzuki}}]{Kunitomo2021}
{Kunitomo}, M., {Ida}, S., {Takeuchi}, T., {et~al.} 2021, \apj, 909, 109

\bibitem[{{Kunitomo} {et~al.}(2020){Kunitomo}, {Suzuki}, \&
  {Inutsuka}}]{Kunitomo2020}
{Kunitomo}, M., {Suzuki}, T.~K., \& {Inutsuka}, S.-i. 2020, \mnras, 492, 3849

\bibitem[{{Lagos} {et~al.}(2021){Lagos}, {Schreiber}, {Zorotovic},
  {G{\"a}nsicke}, {Ronco}, \& {Hamers}}]{Lagos2021}
{Lagos}, F., {Schreiber}, M.~R., {Zorotovic}, M., {et~al.} 2021, \mnras, 501,
  676

\bibitem[{{Lagos} {et~al.}(2023){Lagos}, {Zorotovic}, {Schreiber}, \&
  {G{\"a}nsicke}}]{Lagos2023}
{Lagos}, F., {Zorotovic}, M., {Schreiber}, M.~R., \& {G{\"a}nsicke}, B.~T.
  2023, \mnras, 519, 2302

\bibitem[{{Lagrange} {et~al.}(2010){Lagrange}, {Bonnefoy}, {Chauvin}, {Apai},
  {Ehrenreich}, {Boccaletti}, {Gratadour}, {Rouan}, {Mouillet}, {Lacour}, \&
  {Kasper}}]{Lagrange2010}
{Lagrange}, A.~M., {Bonnefoy}, M., {Chauvin}, G., {et~al.} 2010, Science, 329,
  57

\bibitem[{{Lagrange} {et~al.}(2009){Lagrange}, {Gratadour}, {Chauvin}, {Fusco},
  {Ehrenreich}, {Mouillet}, {Rousset}, {Rouan}, {Allard}, {Gendron}, {Charton},
  {Mugnier}, {Rabou}, {Montri}, \& {Lacombe}}]{Lagrange2009}
{Lagrange}, A.~M., {Gratadour}, D., {Chauvin}, G., {et~al.} 2009, \aap, 493,
  L21

\bibitem[{{Lagrange} {et~al.}(2019){Lagrange}, {Meunier}, {Rubini}, {Keppler},
  {Galland}, {Chapellier}, {Michel}, {Balona}, {Beust}, {Guillot}, {Grandjean},
  {Borgniet}, {M{\'e}karnia}, {Wilson}, {Kiefer}, {Bonnefoy}, {Lillo-Box},
  {Pantoja}, {Jones}, {Iglesias}, {Rodet}, {Diaz}, {Zapata}, {Abe}, \&
  {Schmider}}]{Lagrange2019}
{Lagrange}, A.~M., {Meunier}, N., {Rubini}, P., {et~al.} 2019, Nature
  Astronomy, 3, 1135

\bibitem[{{Lambrechts} \& {Johansen}(2012)}]{Lambrechts12}
{Lambrechts}, M. \& {Johansen}, A. 2012, \aap, 544, A32

\bibitem[{{Lambrechts} {et~al.}(2014){Lambrechts}, {Johansen}, \&
  {Morbidelli}}]{Lambrechts14}
{Lambrechts}, M., {Johansen}, A., \& {Morbidelli}, A. 2014, \aap, 572, A35

\bibitem[{{Lambrechts} {et~al.}(2019){Lambrechts}, {Morbidelli}, {Jacobson},
  {Johansen}, {Bitsch}, {Izidoro}, \& {Raymond}}]{Lambrechts19}
{Lambrechts}, M., {Morbidelli}, A., {Jacobson}, S.~A., {et~al.} 2019, \aap,
  627, A83

\bibitem[{{Ledda} {et~al.}(2023){Ledda}, {Danielski}, \&
  {Turrini}}]{2023A&A...675A.184L}
{Ledda}, S., {Danielski}, C., \& {Turrini}, D. 2023, \aap, 675, A184

\bibitem[{{Luhman}(2022)}]{Luhman2022}
{Luhman}, K.~L. 2022, \aj, 163, 25

\bibitem[{{Lynden-Bell} \& {Pringle}(1974)}]{Lynden-Bell1974}
{Lynden-Bell}, D. \& {Pringle}, J.~E. 1974, \mnras, 168, 603

\bibitem[{{Malamud} \& {Perets}(2020)}]{Malamud2020}
{Malamud}, U. \& {Perets}, H.~B. 2020, \mnras, 492, 5561

\bibitem[{{Maldonado} {et~al.}(2021){Maldonado}, {Villaver}, {Mustill},
  {Ch{\'a}vez}, \& {Bertone}}]{Maldonado2021}
{Maldonado}, R.~F., {Villaver}, E., {Mustill}, A.~J., {Ch{\'a}vez}, M., \&
  {Bertone}, E. 2021, \mnras, 501, L43

\bibitem[{{Manser} {et~al.}(2019){Manser}, {G{\"a}nsicke}, {Eggl}, {Hollands},
  {Izquierdo}, {Koester}, {Landstreet}, {Lyra}, {Marsh}, {Meru}, {Mustill},
  {Rodr{\'\i}guez-Gil}, {Toloza}, {Veras}, {Wilson}, {Burleigh}, {Davies},
  {Farihi}, {Gentile Fusillo}, {de Martino}, {Parsons}, {Quirrenbach}, {Raddi},
  {Reffert}, {Del Santo}, {Schreiber}, {Silvotti}, {Toonen}, {Villaver},
  {Wyatt}, {Xu}, \& {Portegies Zwart}}]{Manser2019}
{Manser}, C.~J., {G{\"a}nsicke}, B.~T., {Eggl}, S., {et~al.} 2019, Science,
  364, 66

\bibitem[{{Marigo} {et~al.}(2020){Marigo}, {Cummings}, {Curtis}, {Kalirai},
  {Chen}, {Tremblay}, {Ramirez-Ruiz}, {Bergeron}, {Bladh}, {Bressan},
  {Girardi}, {Pastorelli}, {Trabucchi}, {Cheng}, {Aringer}, \&
  {Tio}}]{2020NatAs...4.1102M}
{Marigo}, P., {Cummings}, J.~D., {Curtis}, J.~L., {et~al.} 2020, Nature
  Astronomy, 4, 1102

\bibitem[{{Marois} {et~al.}(2008){Marois}, {Macintosh}, {Barman}, {Zuckerman},
  {Song}, {Patience}, {Lafreni{\`e}re}, \& {Doyon}}]{Marois2008}
{Marois}, C., {Macintosh}, B., {Barman}, T., {et~al.} 2008, Science, 322, 1348

\bibitem[{{Marois} {et~al.}(2010){Marois}, {Zuckerman}, {Konopacky},
  {Macintosh}, \& {Barman}}]{Marois2010}
{Marois}, C., {Zuckerman}, B., {Konopacky}, Q.~M., {Macintosh}, B., \&
  {Barman}, T. 2010, \nat, 468, 1080

\bibitem[{{Masset} {et~al.}(2006){Masset}, {Morbidelli}, {Crida}, \&
  {Ferreira}}]{Masset2006}
{Masset}, F.~S., {Morbidelli}, A., {Crida}, A., \& {Ferreira}, J. 2006, \apj,
  642, 478

\bibitem[{{Migaszewski}(2015)}]{Migaszewski2015}
{Migaszewski}, C. 2015, \mnras, 453, 1632

\bibitem[{{Millan-Gabet} {et~al.}(2007){Millan-Gabet}, {Malbet}, {Akeson},
  {Leinert}, {Monnier}, \& {Waters}}]{Millan2007}
{Millan-Gabet}, R., {Malbet}, F., {Akeson}, R., {et~al.} 2007, in Protostars
  and Planets V, ed. B.~{Reipurth}, D.~{Jewitt}, \& K.~{Keil}, 539

\bibitem[{{Monsch} {et~al.}(2021){Monsch}, {Picogna}, {Ercolano}, \&
  {Preibisch}}]{Monsch2021}
{Monsch}, K., {Picogna}, G., {Ercolano}, B., \& {Preibisch}, T. 2021, \aap,
  650, A199

\bibitem[{{Mordasini} {et~al.}(2009){Mordasini}, {Alibert}, {Benz}, \&
  {Naef}}]{Mordasini2009}
{Mordasini}, C., {Alibert}, Y., {Benz}, W., \& {Naef}, D. 2009, \aap, 501, 1161

\bibitem[{{Mu{\~n}oz} \& {Petrovich}(2020)}]{Munoz2021}
{Mu{\~n}oz}, D.~J. \& {Petrovich}, C. 2020, \apjl, 904, L3

\bibitem[{{Mustill} \& {Villaver}(2012)}]{Mustill2012}
{Mustill}, A.~J. \& {Villaver}, E. 2012, \apj, 761, 121

\bibitem[{{Nakatani} {et~al.}(2018){Nakatani}, {Hosokawa}, {Yoshida}, {Nomura},
  \& {Kuiper}}]{Nakatani2018a}
{Nakatani}, R., {Hosokawa}, T., {Yoshida}, N., {Nomura}, H., \& {Kuiper}, R.
  2018, \apj, 865, 75

\bibitem[{{Niedzielski} {et~al.}(2016{\natexlab{a}}){Niedzielski},
  {Deka-Szymankiewicz}, {Adamczyk}, {Adam{\'o}w}, {Nowak}, \&
  {Wolszczan}}]{Nied2016a}
{Niedzielski}, A., {Deka-Szymankiewicz}, B., {Adamczyk}, M., {et~al.}
  2016{\natexlab{a}}, \aap, 585, A73

\bibitem[{{Niedzielski} {et~al.}(2016{\natexlab{b}}){Niedzielski}, {Villaver},
  {Nowak}, {Adam{\'o}w}, {Maciejewski}, {Kowalik}, {Wolszczan},
  {Deka-Szymankiewicz}, \& {Adamczyk}}]{Nied2016b}
{Niedzielski}, A., {Villaver}, E., {Nowak}, G., {et~al.} 2016{\natexlab{b}},
  \aap, 589, L1

\bibitem[{{O'Connor} {et~al.}(2021){O'Connor}, {Liu}, \& {Lai}}]{Oconnor2021}
{O'Connor}, C.~E., {Liu}, B., \& {Lai}, D. 2021, \mnras, 501, 507

\bibitem[{{Ormel} {et~al.}(2010){Ormel}, {Dullemond}, \& {Spaans}}]{Ormel10}
{Ormel}, C.~W., {Dullemond}, C.~P., \& {Spaans}, M. 2010, \apjl, 714, L103

\bibitem[{{Ormel} {et~al.}(2021){Ormel}, {Vazan}, \& {Brouwers}}]{Ormel2021}
{Ormel}, C.~W., {Vazan}, A., \& {Brouwers}, M.~G. 2021, \aap, 647, A175

\bibitem[{{Owen} {et~al.}(2012){Owen}, {Clarke}, \& {Ercolano}}]{Owen2012}
{Owen}, J.~E., {Clarke}, C.~J., \& {Ercolano}, B. 2012, \mnras, 422, 1880

\bibitem[{{Owen} {et~al.}(2010){Owen}, {Ercolano}, {Clarke}, \& {Alexand
  er}}]{Owen2010}
{Owen}, J.~E., {Ercolano}, B., {Clarke}, C.~J., \& {Alexand er}, R.~D. 2010,
  \mnras, 401, 1415

\bibitem[{{Paxton} {et~al.}(2011){Paxton}, {Bildsten}, {Dotter}, {Herwig},
  {Lesaffre}, \& {Timmes}}]{Paxton2011}
{Paxton}, B., {Bildsten}, L., {Dotter}, A., {et~al.} 2011, \apjs, 192, 3

\bibitem[{{Perets}(2010)}]{2010arXiv1001.0581P}
{Perets}, H.~B. 2010, arXiv e-prints, arXiv:1001.0581

\bibitem[{{Picogna} {et~al.}(2021){Picogna}, {Ercolano}, \&
  {Espaillat}}]{Picogna2021}
{Picogna}, G., {Ercolano}, B., \& {Espaillat}, C.~C. 2021, \mnras, 508, 3611

\bibitem[{{Picogna} {et~al.}(2019){Picogna}, {Ercolano}, {Owen}, \&
  {Weber}}]{Picogna2019}
{Picogna}, G., {Ercolano}, B., {Owen}, J.~E., \& {Weber}, M.~L. 2019, \mnras,
  487, 691

\bibitem[{{Pinilla} {et~al.}(2013){Pinilla}, {Birnstiel}, {Benisty}, {Ricci},
  {Natta}, {Dullemond}, {Dominik}, \& {Testi}}]{Pinilla2013}
{Pinilla}, P., {Birnstiel}, T., {Benisty}, M., {et~al.} 2013, \aap, 554, A95

\bibitem[{{Pinilla} {et~al.}(2022){Pinilla}, {Garufi}, \&
  {G{\'a}rate}}]{Pinilla2022}
{Pinilla}, P., {Garufi}, A., \& {G{\'a}rate}, M. 2022, \aap, 662, L8

\bibitem[{{Pollack} {et~al.}(1996){Pollack}, {Hubickyj}, {Bodenheimer},
  {Lissauer}, {Podolak}, \& {Greenzweig}}]{Pollack1996}
{Pollack}, J.~B., {Hubickyj}, O., {Bodenheimer}, P., {et~al.} 1996, \icarus,
  124, 62

\bibitem[{{Pringle}(1981)}]{Pringle1981}
{Pringle}, J.~E. 1981, \araa, 19, 137

\bibitem[{{Reffert} {et~al.}(2015){Reffert}, {Bergmann}, {Quirrenbach},
  {Trifonov}, \& {K{\"u}nstler}}]{Reffert2015}
{Reffert}, S., {Bergmann}, C., {Quirrenbach}, A., {Trifonov}, T., \&
  {K{\"u}nstler}, A. 2015, \aap, 574, A116

\bibitem[{{Ribas} {et~al.}(2015){Ribas}, {Bouy}, \& {Mer{\'{\i}}n}}]{Ribas15}
{Ribas}, {\'A}., {Bouy}, H., \& {Mer{\'{\i}}n}, B. 2015, \aap, 576, A52

\bibitem[{{Richling} \& {Yorke}(1998)}]{RichlingYorke1998}
{Richling}, S. \& {Yorke}, H.~W. 1998, \aap, 340, 508

\bibitem[{{Ronco} {et~al.}(2021){Ronco}, {Guilera}, {Cuadra}, {Miller
  Bertolami}, {Cuello}, {Fontecilla}, {Poblete}, \& {Bayo}}]{Ronco2021}
{Ronco}, M.~P., {Guilera}, O.~M., {Cuadra}, J., {et~al.} 2021, \apj, 916, 113

\bibitem[{{Ronco} {et~al.}(2017){Ronco}, {Guilera}, \& {de
  El{\'\i}a}}]{Ronco2017}
{Ronco}, M.~P., {Guilera}, O.~M., \& {de El{\'\i}a}, G.~C. 2017, \mnras, 471,
  2753

\bibitem[{{Ronco} {et~al.}(2020){Ronco}, {Schreiber}, {Giuppone}, {Veras},
  {Cuadra}, \& {Guilera}}]{Ronco2020}
{Ronco}, M.~P., {Schreiber}, M.~R., {Giuppone}, C.~A., {et~al.} 2020, \apjl,
  898, L23

\bibitem[{{Sellek} {et~al.}(2022){Sellek}, {Clarke}, \&
  {Ercolano}}]{Sellek2022}
{Sellek}, A.~D., {Clarke}, C.~J., \& {Ercolano}, B. 2022, \mnras, 514, 535

\bibitem[{{Shakura} \& {Sunyaev}(1973)}]{ShakuraSunyaev1973}
{Shakura}, N.~I. \& {Sunyaev}, R.~A. 1973, \aap, 24, 337

\bibitem[{{Stahler} \& {Palla}(2004)}]{StahlerPalla2004}
{Stahler}, S.~W. \& {Palla}, F. 2004, {The Formation of Stars}

\bibitem[{{Stapper} {et~al.}(2022){Stapper}, {Hogerheijde}, {van Dishoeck}, \&
  {Mentel}}]{Stapper2022}
{Stapper}, L.~M., {Hogerheijde}, M.~R., {van Dishoeck}, E.~F., \& {Mentel}, R.
  2022, \aap, 667, C1

\bibitem[{{Stephan} {et~al.}(2021){Stephan}, {Naoz}, \& {Gaudi}}]{Stephan2021}
{Stephan}, A.~P., {Naoz}, S., \& {Gaudi}, B.~S. 2021, \apj, 922, 4

\bibitem[{{Suzuki} {et~al.}(2016){Suzuki}, {Ogihara}, {Morbidelli}, {Crida}, \&
  {Guillot}}]{Suzuki2016}
{Suzuki}, T.~K., {Ogihara}, M., {Morbidelli}, A., {Crida}, A., \& {Guillot}, T.
  2016, \aap, 596, A74

\bibitem[{{Tremblay} {et~al.}(2016){Tremblay}, {Cummings}, {Kalirai},
  {G{\"a}nsicke}, {Gentile-Fusillo}, \& {Raddi}}]{Tremblay2016}
{Tremblay}, P.~E., {Cummings}, J., {Kalirai}, J.~S., {et~al.} 2016, \mnras,
  461, 2100

\bibitem[{{van der Marel} \& {Mulders}(2021)}]{vanderMarel2021}
{van der Marel}, N. \& {Mulders}, G.~D. 2021, \aj, 162, 28

\bibitem[{{Vanderburg} {et~al.}(2015){Vanderburg}, {Johnson}, {Rappaport},
  {Bieryla}, {Irwin}, {Lewis}, {Kipping}, {Brown}, {Dufour}, {Ciardi}, {Angus},
  {Schaefer}, {Latham}, {Charbonneau}, {Beichman}, {Eastman}, {McCrady},
  {Wittenmyer}, \& {Wright}}]{Vandenburg2015}
{Vanderburg}, A., {Johnson}, J.~A., {Rappaport}, S., {et~al.} 2015, \nat, 526,
  546

\bibitem[{{Vanderburg} {et~al.}(2020){Vanderburg}, {Rappaport}, {Xu},
  {Crossfield}, {Becker}, {Gary}, {Murgas}, {Blouin}, {Kaye}, {Palle}, {Melis},
  {Morris}, {Kreidberg}, {Gorjian}, {Morley}, {Mann}, {Parviainen}, {Pearce},
  {Newton}, {Carrillo}, {Zuckerman}, {Nelson}, {Zeimann}, {Brown},
  {Tronsgaard}, {Klein}, {Ricker}, {Vanderspek}, {Latham}, {Seager}, {Winn},
  {Jenkins}, {Adams}, {Benneke}, {Berardo}, {Buchhave}, {Caldwell},
  {Christiansen}, {Collins}, {Col{\'o}n}, {Daylan}, {Doty}, {Doyle},
  {Dragomir}, {Dressing}, {Dufour}, {Fukui}, {Glidden}, {Guerrero}, {Guo},
  {Heng}, {Henriksen}, {Huang}, {Kaltenegger}, {Kane}, {Lewis}, {Lissauer},
  {Morales}, {Narita}, {Pepper}, {Rose}, {Smith}, {Stassun}, \&
  {Yu}}]{Vandenburg2020}
{Vanderburg}, A., {Rappaport}, S.~A., {Xu}, S., {et~al.} 2020, \nat, 585, 363

\bibitem[{{Venturini} {et~al.}(2016){Venturini}, {Alibert}, \&
  {Benz}}]{Venturini16}
{Venturini}, J., {Alibert}, Y., \& {Benz}, W. 2016, \aap, 596, A90

\bibitem[{{Venturini} {et~al.}(2020{\natexlab{a}}){Venturini}, {Guilera},
  {Haldemann}, {Ronco}, \& {Mordasini}}]{Venturini2020Letter}
{Venturini}, J., {Guilera}, O.~M., {Haldemann}, J., {Ronco}, M.~P., \&
  {Mordasini}, C. 2020{\natexlab{a}}, \aap, 643, L1

\bibitem[{{Venturini} {et~al.}(2020{\natexlab{b}}){Venturini}, {Guilera},
  {Ronco}, \& {Mordasini}}]{Venturini2020SE}
{Venturini}, J., {Guilera}, O.~M., {Ronco}, M.~P., \& {Mordasini}, C.
  2020{\natexlab{b}}, \aap, 644, A174

\bibitem[{{Venturini} {et~al.}(2020{\natexlab{c}}){Venturini}, {Ronco}, \&
  {Guilera}}]{Venturini2020Rev}
{Venturini}, J., {Ronco}, M.~P., \& {Guilera}, O.~M. 2020{\natexlab{c}}, \ssr,
  216, 86

\bibitem[{{Veras}(2021)}]{Veras2021}
{Veras}, D. 2021, in Oxford Research Encyclopedia of Planetary Science, 1

\bibitem[{{Veras} {et~al.}(2014){Veras}, {Leinhardt}, {Bonsor}, \&
  {G{\"a}nsicke}}]{Veras2014}
{Veras}, D., {Leinhardt}, Z.~M., {Bonsor}, A., \& {G{\"a}nsicke}, B.~T. 2014,
  \mnras, 445, 2244

\bibitem[{{Wang} \& {Goodman}(2017)}]{WangGoodman2017}
{Wang}, L. \& {Goodman}, J. 2017, \apj, 847, 11

\bibitem[{{Williams} \& {Cieza}(2011)}]{WilliamsCieza2011}
{Williams}, J.~P. \& {Cieza}, L.~A. 2011, \araa, 49, 67

\bibitem[{{Wolthoff} {et~al.}(2022){Wolthoff}, {Reffert}, {Quirrenbach},
  {Jones}, {Wittenmyer}, \& {Jenkins}}]{Wolthoff2022}
{Wolthoff}, V., {Reffert}, S., {Quirrenbach}, A., {et~al.} 2022, \aap, 661, A63

\bibitem[{{Yasui} {et~al.}(2014){Yasui}, {Kobayashi}, {Tokunaga}, \&
  {Saito}}]{Yasui2014}
{Yasui}, C., {Kobayashi}, N., {Tokunaga}, A.~T., \& {Saito}, M. 2014, \mnras,
  442, 2543

\bibitem[{{Zhu} \& {Dong}(2021)}]{Zhu2021}
{Zhu}, W. \& {Dong}, S. 2021, \araa, 59, 291

\end{thebibliography}

\begin{appendix} 
\section{Code validation and comparison with \citet{Kunitomo2021}}\label{AppA}

\begin{figure*}
   \centering
   \includegraphics[angle=-90,width=0.90\linewidth]{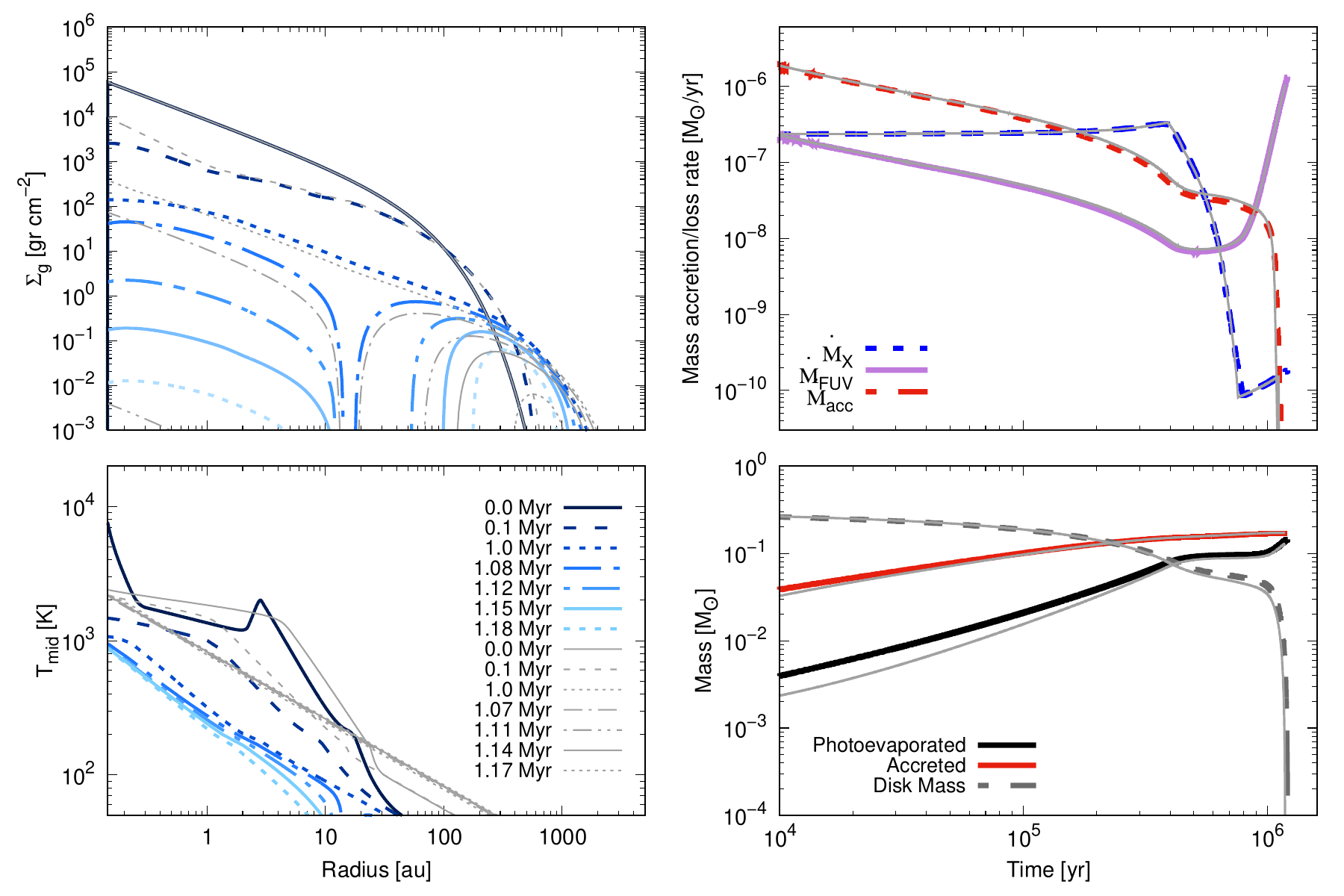}
   \caption{Time evolution of a disc around a $3M_\odot$ star computed with this paper version of {\scriptsize PLANETALP}. As in fig. 8 of \citet{Kunitomo2021}, panels on the left show the time evolution of the gas surface density (top) and the midplane temperature (bottom) profiles. Panels on the right shows the time evolution of the mass accretion rate $\dot{M}_{\text{acc}}$ and
the mass-loss rates by the X-ray and FUV photoevaporation (top), and the evolution of the disc mass and the time-integrated masses of accretion and photoevaporation (FUV and X-ray). The thick grey curves in each panel show the results from \citetalias{Kunitomo2021}}.
   \label{Fig:AppA1}
    \end{figure*}

\begin{figure*}
   \centering
   \includegraphics[angle=-90,width=0.90\linewidth]{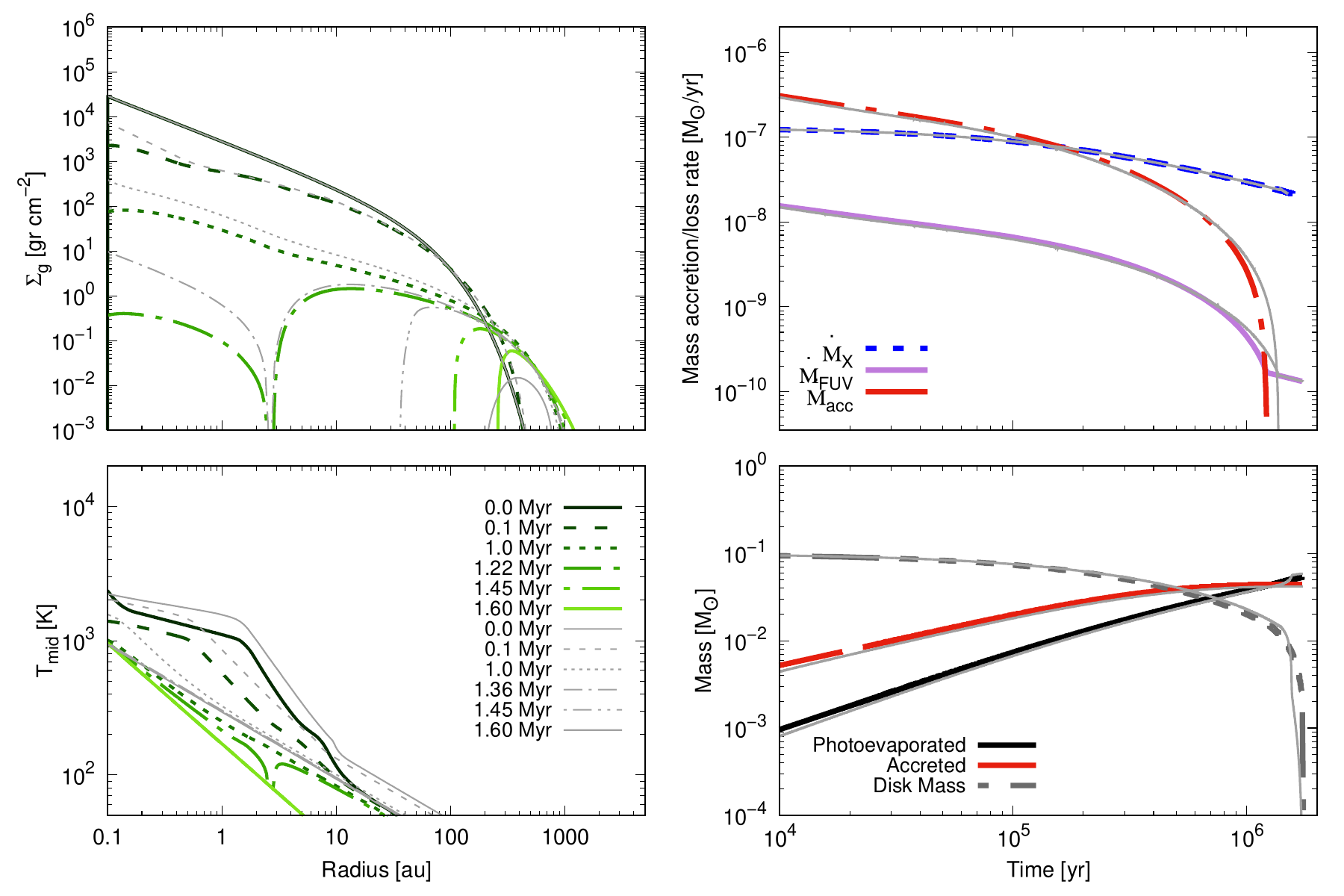}
   \caption{Same as fig. \ref{Fig:AppA1} but for a $1M_\odot$ star. This figure compares to fig. 15 of \citetalias{Kunitomo2021}.}
   \label{Fig:AppA2}
    \end{figure*}

In order to validate our 1D+1D model for the evolution of the gas component in protoplanetary discs considering the impact of the stellar evolution on the X-ray and FUV photoevaporation models (added in {\scriptsize PLANETALP}), we compare our results against those of \citetalias{Kunitomo2021}, for their cases with $\Mstar=3\Msun$ (their figure 8) and $\Mstar=1\Msun$ (their figure 15). To do this we consider the parameters for their fiducial disc model with $\alpha_{0}= 10^{-2}$(see their Table 2). We follow their disc dispersal criterion, this is the disc completely dissipates when its mass decreases down to $10^{-8}{\text{M}_{\text{d}}}$. Our grid also uses 2000 radial bins, in our case logarithmically equally spaced, between 0.1 au and $10^4$ au for a $1\Msun$ star, and between 0.144 au $10^4$ au for a $3\Msun$ star. Despite in \citetalias{Kunitomo2021} the calculation domain ranges from 0.01 au, our results do not present differences due to this choice, which was taken to avoid numerical problems in the computation of the vertical structure due to the high disc temperatures obtained very close to the central star, and as a consequence of considering eq. \ref{eq:inner-radius}. We also consider the same boundary condition of zero-torque imposed at both the inner
and outer boundaries of the disc, and we follow the same disc initial profile. Since the authors showed EUV photoevaporation does not play an important role for stars with masses between $1\Msun$ and $3\Msun$, we did not include it in our model. 
The main difference between our work and the one by \citetalias{Kunitomo2021} is that we compute in detail the vertical structure of the disc (see \citet{Guilera2017Letter}). This affects and changes the evolution of the disc temperature profiles and, as a consequence, changes the way the gas surface density evolves (see left panels of figs. \ref{Fig:AppA1} and \ref{Fig:AppA2} where the colored curves represent our results and the thick grey curves correspond to \citetalias{Kunitomo2021}'s results). However, as it can be seen in the right panels of figs. \ref{Fig:AppA1} and \ref{Fig:AppA2}, the time evolution of the mass accretion and mass loss rates by photoevaporation, the time evolution of the mass of the disc and the time-integrated masses of accretion and photoevaporation are practically the same as those of \citetalias{Kunitomo2021} (represented with grey thick curves). 

In our model, the detailed computation of the vertical structure leads to an evolution of the temperature profiles that presents values between $\sim$2 and several times lower along the disc than those of \citetalias{Kunitomo2021} for the $3M_\odot$ star case. We associate this difference with the variance between the irradiation temperature adopted by \citetalias{Kunitomo2021} which is the classical irradiation temperature from \citet{Hayashi81} (see eq. 6 in \citet{Kunitomo2020}), and the more detailed one adopted by us following \citet{Migaszewski2015} (see eq. \ref{eq:temp_irrad}). Since the disc viscosity depends on the sound speed as $\nu=\alpha c^2_{\text{s}}/\Omega$, with $c^2_{\text{s}}=P/\rho$, and $P$ relates with the temperature $T$ through the equation of state of an ideal diatomic gas $P = \rho \kappa T/\mu m_{\text{H}}$, higher disc temperatures lead to more viscous discs. Thus, despite the use of the same $\alpha$ viscosity parameter, discs in \citetalias{Kunitomo2021} evolve viscously faster than our discs. As a consequence, the gas surface density profiles for a disc around a $3\Msun$ star with $\alpha\propto 10^{-2}$ do not show an 'homogenous evolution' of the inner and outer discs after the gap opening, as we do.

We recall that in our model, the vertical structure of the disc is computed but only once and at the beginning, before solving Eq. \ref{eq:evol_gas}, for fixed values of $\Teff$ and $\Rstar$ that correspond to those at the birthline. This is, we are not considering the temporal change of the stellar radius and effective temperature to compute the irradiation onto the disc's surface by the central star. Despite this simplification, which will be explicitly address in the next section, we do not find differences with the results of \citetalias{Kunitomo2021}.  

\section{Changes in the vertical structure with \Teff and \Rstar}\label{AppB}

As it was mentioned in the previous section, our model computes the vertical structure without taking into account the temporal change of the stellar radius and the effective temperature in the disc's surface irradiation by the central star. And since we do not find significant differences between our results and those of \citetalias{Kunitomo2021} we can say that this is a good approximation. However, one could argue that this is perhaps due to the high viscosity considered in the previous disc simulations ($\alpha \propto 10^{-2}$), which could cause that the heat generated by viscosity is much more relevant than the one generated by the stellar irradiation itself. To show this is not the case and that our approximation is valid in general, we here compute the time evolution of the same disc around a $3M_\odot$ star as before, but with a much lower viscosity parameter $\alpha=3\times10^{-4}$, and with the vertical structure computed from values of $T_\star^{\text{eff}}$ and $R_\star$ at 0 Myr (\citetalias{Kunitomo2021}'s birthline), 0.5~Myr and 1.0~Myr, this is once the changes in $T_\star^{\text{eff}}$, $R_\star$ and $L_\star$ become relevant for a $3M_\odot$ star (see Table 1 in \citetalias{Kunitomo2021}). 

\begin{figure}
   \centering
   \includegraphics[angle= 270, width=0.95\columnwidth]{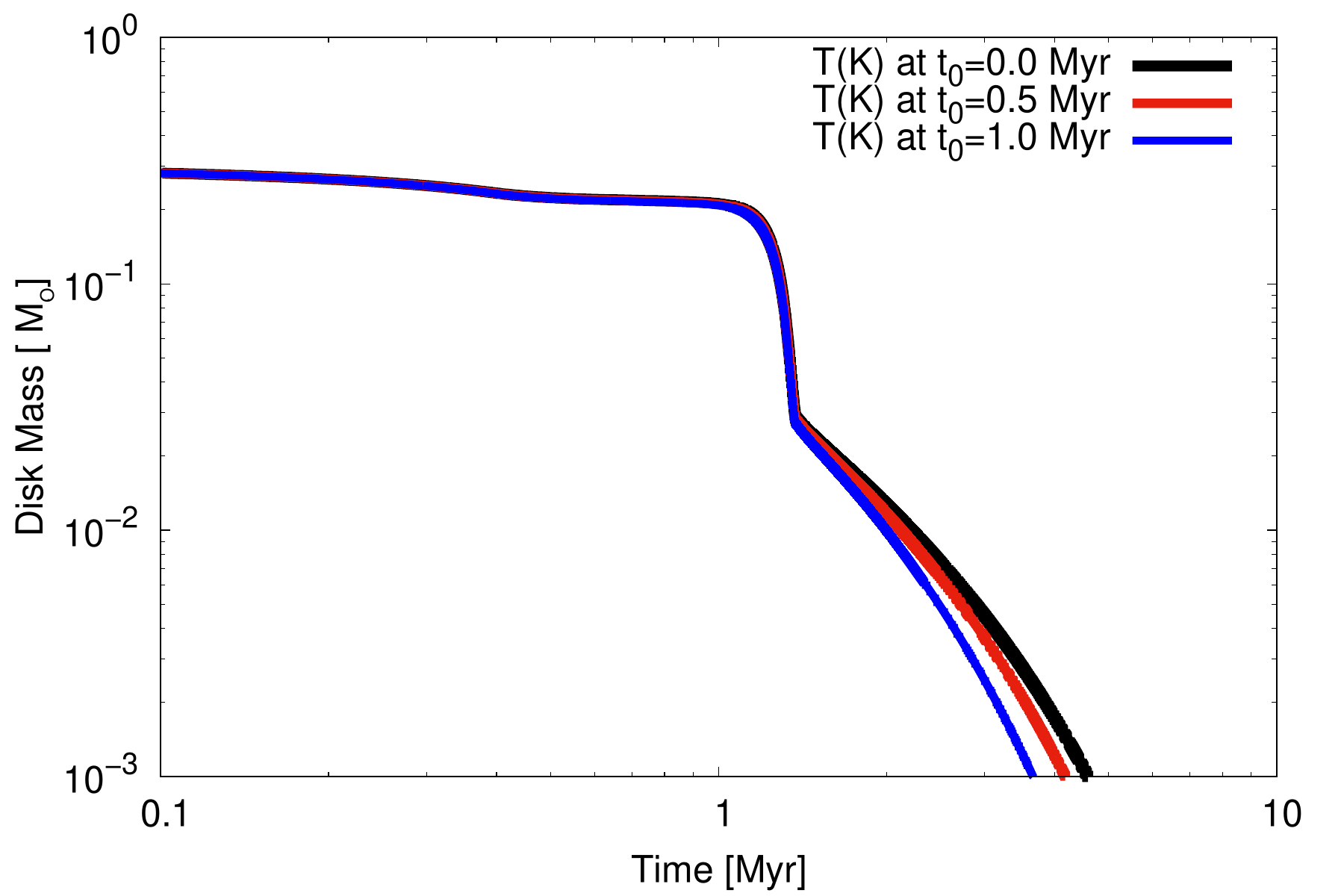}
   \caption{Time evolution of the mass of a disc around a $3M_\odot$ starwith its vertical structure computed from different temperatures and star radius (at $t_0=0$, $t_0=0.5$ and $t_0=1.0$ Myr (data from Table 1 in \citetalias{Kunitomo2021}).}
   \label{Fig:AppB}
    \end{figure}

In figure \ref{Fig:AppB} we show the time evolution of the disc mass for each case. During the first 1.37 Myr, which is the time during which the FUV photoevaporation efficiently acts by completely removing the outer disc (after the gap opening), the three discs evolve exactly the same and on the same time scale. Once the outer disc has been removed, only the internal disc, which loses mass only by viscous accretion (similar to the evolution of the S45 disc in fig. \ref{Fig:Perfiles-y-Mpunto-Alpha4}), remains. Only at this stage we find little differences between their final dissipation timescales. However these little differences do not change significantly the total disc lifetimes nor the evolutionary pathways they present.

\longtab[1]{
\begin{landscape}
\begin{longtable}{lrcrrrrrrrrl}
...
\end{longtable}
\end{landscape}
}
\end{appendix}
\end{document}